\documentclass[aip,jap,reprint,longbibliography]{revtex4-1}

\usepackage{graphicx}
\usepackage{dcolumn}
\usepackage{bm}

\usepackage[utf8]{inputenc}
\usepackage[T1]{fontenc}
\usepackage{mathptmx}

\usepackage{subfigure}
\usepackage{amsfonts}
\usepackage{amssymb}
\usepackage{amsmath}
\usepackage{color}
\usepackage{multirow}
\usepackage[colorlinks=true,citecolor=blue,urlcolor=blue,linkcolor=blue,hyperfigures=true]{hyperref}
\usepackage{comment}

\graphicspath{{./Figures/}}

\def\Eq{Eq.~}
\def\Eqs{Eqs.~}
\def\Fig{Fig.~}
\def\Figs{Figs.~}
\def\Ref{Ref.~}
\def\Refs{Refs.~}
\def\Sec{Sec.~}

\def\be{\begin{equation}}
\def\ee{\end{equation}}
\def\bea{\begin{eqnarray}}
\def\eea{\end{eqnarray}}

\def\ie{\textit{i.e.}~}
\def\eg{\textit{e.g.}~}
\def\etal{\textit{et al}~}
\def\dd{\mathrm{d}}
\def\ud{\uparrow\downarrow}

\newcommand{\ket}[1]{\left| #1 \right\rangle}

\newcommand{\refeq}[1]{(\ref{#1})}

\begin{document}


\title[Testing the UFF using correlated $^{39}$K -- $^{87}$Rb interferometers]{Testing the Universality of Free Fall using Correlated $^{39}$K -- $^{87}$Rb Atom Interferometers}

\author{B. Barrett}
\affiliation{LP2N, IOGS, CNRS and Universit\'{e} de Bordeaux, rue Fran\c{c}ois Mitterrand, 33400 Talence France}
\affiliation{Department of Physics, University of New Brunswick, 8 Bailey Drive, Fredericton, NB E3B 5A3, Canada}

\author{G. Condon}
\altaffiliation[Also at ]{Muquans, Institut d’Optique d’Aquitaine,  rue Fran\c{c}ois Mitterrand, 33400 Talence, France}

\author{L. Chichet}
\altaffiliation[Also at ]{Teledyne e2v, 106 Waterhouse Ln, Chelmsford CM1 2QU, United Kingdom}

\author{L. Antoni-Micollier}
\altaffiliation[Also at ]{Muquans, Institut d’Optique d’Aquitaine,  rue Fran\c{c}ois Mitterrand, 33400 Talence, France}

\author{R. Arguel}
\author{M. Rabault}
\altaffiliation[Also at ]{Muquans, Institut d’Optique d’Aquitaine,  rue Fran\c{c}ois Mitterrand, 33400 Talence, France}

\author{C. Pelluet}
\author{V. Jarlaud}

\affiliation{LP2N, IOGS, CNRS and Universit\'{e} de Bordeaux, rue Fran\c{c}ois Mitterrand, 33400 Talence France}

\author{A. Landragin}

\affiliation{LNE-SYRTE, Observatoire de Paris, Universit\'{e} PSL, CNRS, Sorbonne Universit\'{e}, \\
61 avenue de l'Observatoire, 75014 Paris, France}

\author{P. Bouyer}
\homepage{https://www.coldatomsbordeaux.org/}

\author{B. Battelier}
\email{baptiste.battelier@institutoptique.fr}

\affiliation{LP2N, IOGS, CNRS and Universit\'{e} de Bordeaux, rue Fran\c{c}ois Mitterrand, 33400 Talence France}

\date{\today}

\begin{abstract}
We demonstrate how simultaneously-operated $^{39}$K -- $^{87}$Rb interferometers exhibiting a high level of correlation can be used to make competitive tests of the university of free fall. This work provides an overview of our experimental apparatus and data analysis procedure, including a detailed study of systematic effects. With a total interrogation time of $2T = 40$ ms in a compact apparatus, we reach a statistical uncertainty on the measurement of the E\"{o}tv\"{o}s parameter of $7.8 \times 10^{-8}$ after $2.4 \times 10^4$ s of integration. The main limitations to our measurement arise from a combination of wavefront aberrations, the quadratic Zeeman effect in $^{39}$K, parasitic interferometers in $^{87}$Rb, and the velocity sensitivity of our detection system. These systematic errors limit the accuracy of our measurement to $\eta = 0.9(1.6) \times 10^{-6}$. We discuss prospects for improvements using ultracold atoms at extended interrogation times.
\end{abstract}

\maketitle

\section{Introduction}


Einstein's theory of General Relativity (GR) is a cornerstone of our current description of the physical universe at macroscopic scales, and provides our most complete description for the laws of gravitation. Gravity interacts in the same way with any mass/energy, which allows for a geometric description of gravitation as the effect of space-time curvature. The phenomenological manifestation of this universal coupling is known as Einstein's Equivalence Principle (EEP), which hypothesizes an exact correspondence between the gravitational and inertial mass of any object. Precise tests of this principle can advance the search for physics beyond the Standard Model and GR, and may shed new light on our understanding of the universe and its constituents, especially cold dark matter and dark energy \cite{Damour2002, Will2006, Damour2012, Hohensee2013}.

One aspect of the EEP---known as the universality of free fall (UFF)---states that if any uncharged test body is placed at an initial position in space-time and given an initial velocity, then its subsequent trajectory will be independent of its internal structure and composition. Tests of the UFF generally involve measuring the relative acceleration between two different test bodies undergoing free fall within the same gravitational field. Such tests are characterized by the E\"{o}tv\"{o}s parameter:
\be
  \label{eta}
  \eta = 2 \frac{a_1 - a_2}{a_1 + a_2},
\ee
where $a_{i}$ is the gravitational acceleration of object $i$ with respect to the source mass. If the UFF is satisfied, the E\"{o}tv\"{o}s parameter is zero. The UFF has been tested with an uncertainty at the level of $10^{-13}$ with classical test bodies, as with a torsion balance \cite{Williams2004, Schlamminger2008}, and more recently at $2 \times 10^{-14}$ by comparing two electro-static accelerometers onboard the Microscope satellite \cite{Touboul2017}. So far, no violation of the UFF has been measured.

However, it is entirely possible that a violation of the UFF will manifest as a quantum effect that is not included in GR. This motivates using quantum objects, such as clouds of cold atoms, to test the UFF for violations which classical objects may not be sensitive to. One possibility may stem from the fact that the classical and quantum mechanical descriptions of motion are fundamentally different. Consequently, considering quantum phenomena in the context of gravity poses many conceptual and fundamental questions that may shed new light on this frontier of physics \cite{Battelier2019}.


Atom interferometry offers the possibility of measuring the relative acceleration between two different atoms with a sensitivity that scales as $n k T^2$, where $n$ is the number of photon momentum transferred to the atom, $k$ is the wavevector of the interrogation light (typically $\sim 10^7$ rad/m), and $T$ is the free-fall time between light pulses. Presently, among the most precise atom interferometric test of the UFF between two quantum bodies we can cite a measurement using cold rubidium isotopes with different angular momenta ($^{87}$Rb $\ket{F=1}$ and $^{85}$Rb $\ket{F=2}$) \cite{Zhou2021}, reaching an accuracy of $\delta\eta = 6.7 \times 10^{-10}$ using $n = 4$ photon momenta and a free-fall time of $T = 203$ ms. Moreover, \Ref \citenum{Asenbaum2020} reports a sensitivity of $5.4 \times 10^{-11}/\sqrt{\mathrm{Hz}}$ and an uncertainty of $3.4 \times 10^{-12}$ with $n = 12$ and $T = 955$ ms using the two Rb isotopes in the same internal state.
A more complete account of cold-atom-based UFF tests can be found in \Ref \citenum{Albers2020}.

\begin{figure*}[!tb]
  \centering
  \includegraphics[width=0.80\textwidth]{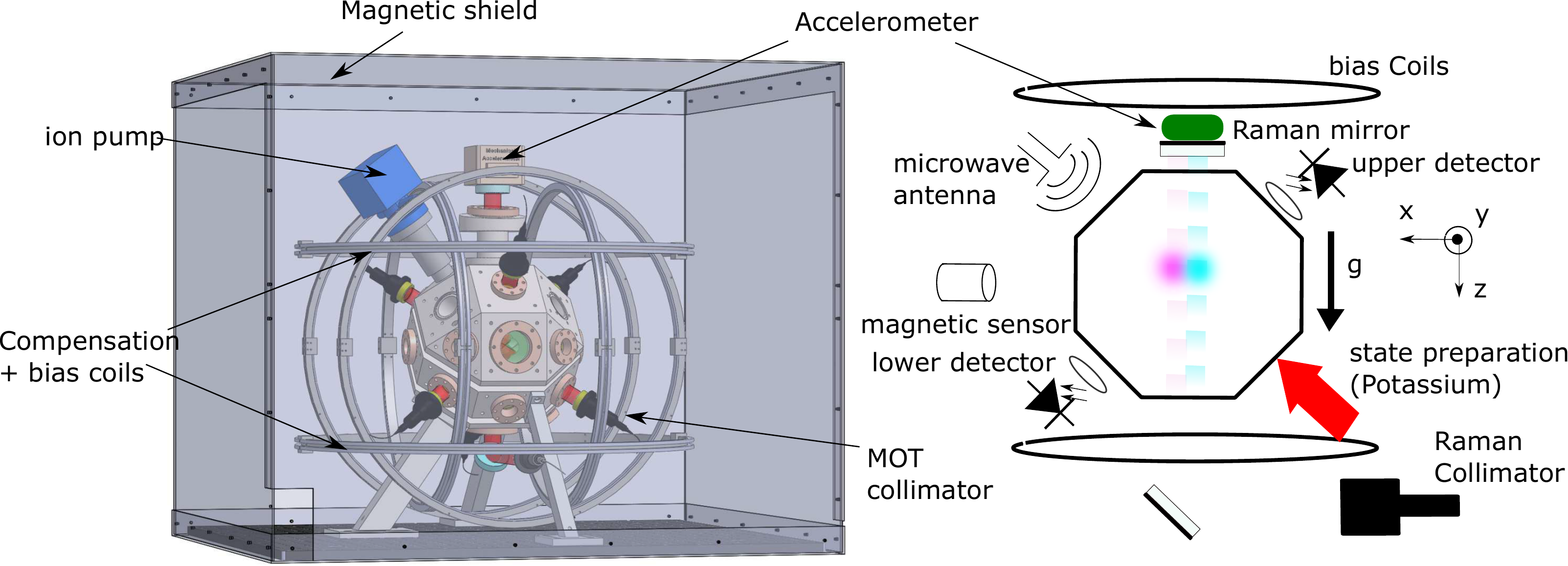}
  \caption{Experimental setup. Left: 3D drawing of our titanium vacuum chamber surrounded by a mu-metal magnetic shield. Right: lateral section of the chamber. Rubidium (pink) and potassium (light blue) are cooled and trapped simultaneously in the central region. Both clouds are released at the same time and their accelerations are compared relative to a common reference frame: a mirror that retro-reflects the two overlapped Raman beams. A large pair of Helmholtz coils on the vertical axis provide a magnetic bias field, and we servo this field using a magnetic sensor located near the atoms. A mechanical accelerometer is fixed to the rear of the reference mirror to correct for vibration-induced phase noise on both interferometers.}
  \label{fig:Experiment}
\end{figure*}

So far, the majority of quantum UFF tests have involved two atoms of the same chemical species (\eg the isotopes of Rb or Sr), and hence they exhibit near-unity mass ratios. A relevant test of gravity theories requires test bodies with very different masses, but for atom interferometers this also implies increased experimental challenges. Nevertheless, three UFF tests using $^{87}$Rb and $^{39}$K have previously been carried out. These atoms exhibit a mass ratio of 2.13 and differ greatly in composition---making them more sensitive to possible UFF violations than isotopic pairs of the same chemical element \cite{Hohensee2013}. Schlippert \etal reported $\delta \eta = 5.4 \times 10^{-7}$ with $n = 2$ and $T = 20$ ms \cite{Schlippert2014}, and the same group recently improved their result to $3.2 \times 10^{-7}$ by increasing the free-fall time to $T = 41$ ms \cite{Albers2020}. Our group previously achieved $3.0 \times 10^{-4}$ with $n = 2$ and $T = 2$ ms in the microgravity environment simulated onboard an aircraft \cite{Barrett2016}. One of the main limitations in each of these experiments is the potassium source, which exhibits larger temperatures, lower atom numbers and lower state purities compared to rubidium. These features decrease the interference contrast and increase systematic effects, which limit both the statistical uncertainty and accuracy of UFF tests.

In this work, we present a new measurement of the E\"{o}tv\"{o}s parameter using highly-correlated $^{87}$Rb and $^{39}$K interferometers. With a modest free-fall time of $T = 20$ ms in a compact apparatus, we reach a statistical uncertainty $\delta \eta^{\rm stat} = 7.8 \times 10^{-8}$ after $2.4 \times 10^4$ s of integration. We evaluate the main systematic effects for each atom interferometer and find $\delta \eta^{\rm sys} = 1.6 \times 10^{-6}$ which is the dominant contribution to our measurement uncertainty. The focus of this work is two-fold. First, we present a detailed analysis of experimental data using two independent methods for correlated dual-species interferometers. One method correlates each atomic sensor with a classical one to reconstruct interference fringes in the absence of a stable inertial frame, while the other relies Bayesian inference using correlations between each atomic sensor output. We show that the two analysis methods are equivalent, while latter shows great promise for future high-precision UFF tests with different chemical species \cite{Hartwig2015}. The second focus of this work is a detailed study of systematic effects with these atoms. We have identified several surprising effects in $^{39}$K which limit our measurement. This study serves as a necessary precursor for future long-baseline interferometry experiments with these species.

The article is organized as follows. Our experimental setup and data acquisition protocol is described in \Sec \ref{sec:Experiment}. In \Sec \ref{sec:Methods}, we present two methods to extract the E\"{o}tv\"{o}s parameter using correlated interferometers and discuss their performance. Section \ref{sec:Systematics} provides a detailed study of systematic effects in our present system. We identify the main limitations and highlight specific challenges working with $^{39}$K. Finally, we conclude and give perspectives for future measurements in \Sec \ref{sec:Conclusion}.

\section{Compact dual-species interferometer}
\label{sec:Experiment}


\subsection{Experimental setup}

Our apparatus was originally designed to operate in a microgravity environment, such as onboard the Novespace Zero-G plane \cite{Barrett2016} or an Einstein elevator \cite{Condon2019}, where large interrogation times will lead to increased measurement sensitivity. However, in standard gravity, the maximum free-fall time is limited to $T \sim 20$ ms due to the close proximity of the detection zone to the chamber center. Here, we provide a brief overview of the apparatus. A more detailed description of the setup can be found in \Refs \citenum{Barrett2015, Barrett2014a, AntoniMicollier2017}.

Our robust and transportable device is comprised of a bank of fiber-based lasers, an ultra-stable frequency source and a titanium science chamber shown in \Fig \ref{fig:Experiment}. The $^{39}$K and $^{87}$Rb atomic sources are derived from overlapped 3D magneto-optical traps (MOTs) loaded from a hot background vapor. Lasers on the D2 transition for both $^{87}$Rb and $^{39}$K (780 and 767 nm, respectively) and one on the D1 transition for $^{39}$K (770 nm) are combined and split equally into six beams. These cooling beams then pass through separate polarization-maintaining fibers to optical collimators attached to the vacuum chamber.

The interferometry beams for $^{87}$Rb ($^{39}$K) are derived from the same laser diodes as the D2 cooling beams via frequency-offset locks at a detuning of $\Delta \simeq -1.1$ GHz ($-1.3$ GHz) relative to the D2 cycling transition $\ket{F = 2} \to \ket{F'=3}$.  The two phase-locked optical frequencies required for Raman transitions are derived from an electo-optic phase modulator operating at 6.8 GHz for $^{87}$Rb, and a dual-pass acousto-optic modulator operating at 230 MHz for $^{39}$K. The absolute frequency stability of both laser systems is $\sim 500$ kHz, corresponding to a relative uncertainty of $\sim 1.5 \times 10^{-9}$ in the wavevector $k$. These beams are combined with the same linear polarization in a commercial collimator, and are aligned through the atoms and a quarter-waveplate along the vertical axis before being retro-reflected by a mirror to create a lin$\perp$lin polarization geometry. This configuration efficiently drives velocity-sensitive two-photon Raman transitions, while suppressing Doppler-insensitive ones.

To prepare the internal state of the $^{87}$Rb sample, we use a 6.8 GHz antenna aligned in the horizontal direction. Similarly, for $^{39}$K we employ a sequence of optical pulses using a dedicated beam aligned at $45^\circ$ relative to the vertical, as shown in \Fig \ref{fig:Experiment}.

The magnetic field at the MOT location is zeroed using three orthogonal pairs of compensation coils, and we correct a small magnetic gradient using a pair of anti-Helmholtz coils along the vertical axis. During the interferometer sequence, we apply a vertical bias field of $\sim 120$ mG. Due to the proximity of the lower coil to the aluminum breadboard supporting the experiment, Eddy currents induce a large temporal variation in the field that decays exponentially over $\sim 50$ ms. We compensate this effect by applying a feedback current on the bias coils during the interferometer and locking the magnetic field measured by a flux-gate sensor (Bartington Instruments MAG03MCTPB500) located near the atoms.

\subsection{Dual-interferometer measurement sequence}

\begin{figure}[!tb]
  \centering
  \includegraphics[width=0.48\textwidth]{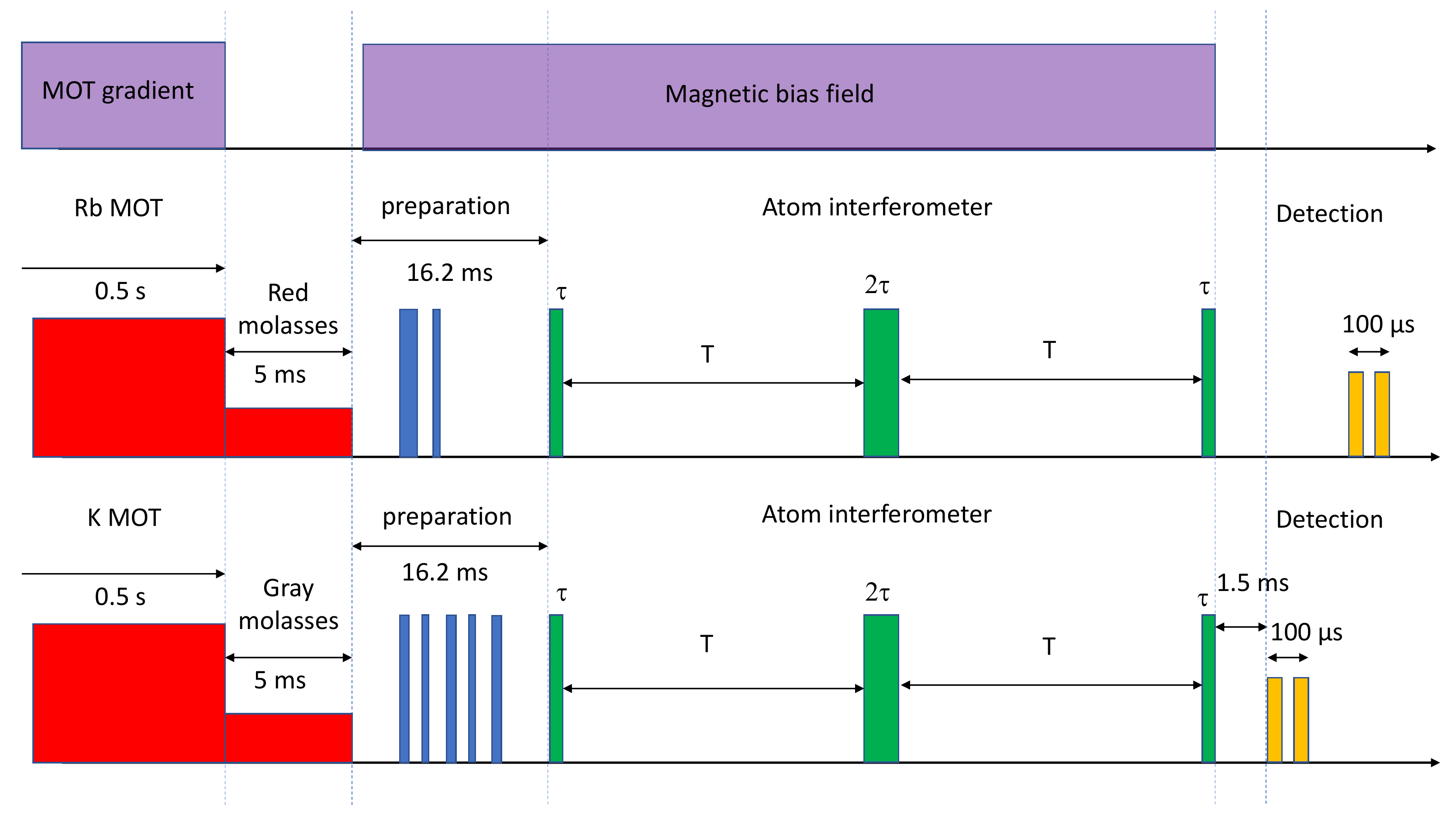}
  \caption{Experimental sequence for the simultaneous K-Rb interferometers ($^{87}$Rb above, $^{39}$K below). The MOT loading and molasses phases for each species are performed simultaneously in order to release the atoms from the same location at the same time. The pulse sequence for the two interferometers is overlapped and identical ($T=20$ ms, $\tau = 2.5$ $\mu$s). The normalized populations in $\ket{F=2}$ are detected in a sequential manner by applying a staggered sequence of resonant pulses (see below).}
  \label{fig:Sequence}
\end{figure}

A dual-species MOT is loaded from background vapour in 0.5 s, containing approximately $6 \times 10^7$ $^{39}$K atoms and $3 \times 10^8$ $^{87}$Rb atoms \footnote{The cycle time for each measurement was $t_{\rm cyc} \simeq 2.1$ s, which includes significant dead time resulting from the software that controls the experiment \cite{Keshet2013}.}. Rubidium atoms are then cooled in a standard red-detuned optical molasses while $^{39}$K atoms are cooled in a gray molasses using blue-detuned light from the D1 line ($\ket{F = 2} \to \ket{F' = 2}$) \cite{Salomon2013}. Both species have a temperature of $\sim 5$ $\mu$K at the end of the molasses cooling stage (see \Fig \ref{fig:Sequence}).

We then prepare both species in the magnetically-insensitive state $\ket{F = 1, m_F = 0}$. A quantization field is applied along the vertical axis in the form of a magnetic bias field of $\sim 120$ mG. For rubidium, we initially repump the atoms to the $\ket{F = 2}$ manifold. We then apply a microwave pulse at 6.834 GHz to transfer atoms into $\ket{F = 1, m_F = 0}$, and a near-resonant push beam subsequently removes atoms remaining in $\ket{F = 2}$. For potassium, we use a state-juggling technique based on a sequence of coherent Raman and optical pumping pulses \cite{AntoniMicollier2017}. After this preparation sequence, the state purity of both species is $> 90\%$ and the temperature of the $^{39}$K sample is $\sim 6$ $\mu$K.

The simultaneous matter-wave interferometers are formed by a $\pi/2-\pi-\pi/2$ sequence of counter-propagating Raman pulses, each separated by interrogation time $T$, with durations $\tau-2\tau-\tau$. This pulse sequence is identical for both species. The bias field is kept on during the interferometer to ensure that $\ket{F = 1, m_F = \pm 1} \to \ket{F = 2, m_F = \pm 1}$ Raman resonances are well-separated from those involving $\ket{F = 1, m_F = 0}$. To compensate for the Doppler shift during free fall, the frequency difference between Raman beams for each species $S$ is linearly chirped, \ie $\delta_{\rm S}(t) = \alpha_{\rm S} t$, such that the chirp rate $\alpha_{\rm S} \simeq k_{\rm S} g$. This phase continuous chirp is also used as a control parameter to scan the interferometer phase and to identify the central fringe, as we discuss below.

\begin{figure}[!tb]
  \centering
  \includegraphics[width=0.48\textwidth]{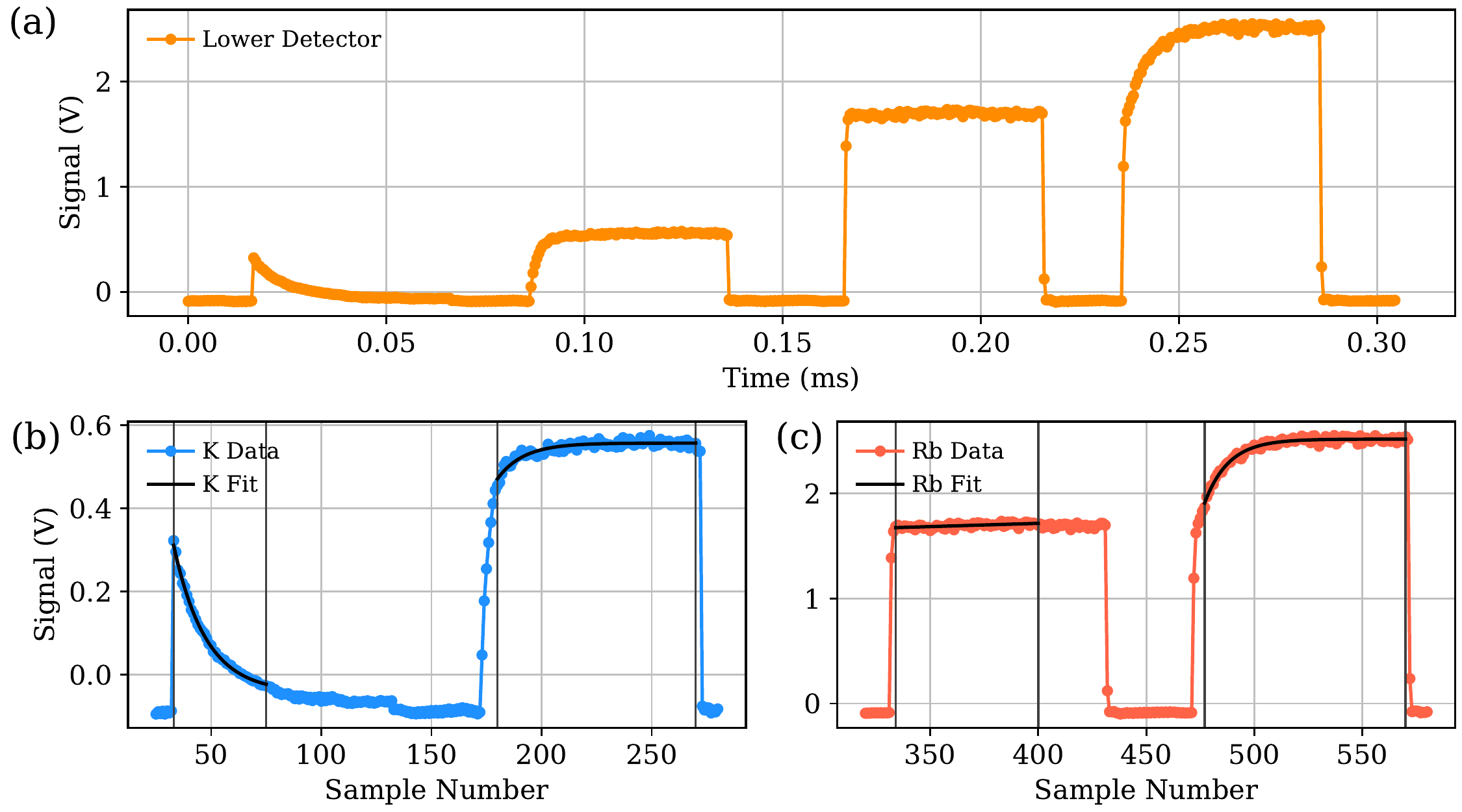}
  \caption{Typical detection signals for $^{39}$K and $^{87}$Rb. (a) Full detection trace recorded by the acquisition system ($\sim 500$ kHz bandwidth). An analysis of $^{39}$K (b) and $^{87}$Rb (c) signals is shown on the bottom row. For each species $S$, fluorescence signals representing $N_{\rm S}^{\ket{2}}$ (left peak) and $N_{\rm S}^{\rm total}$ (right peak) are fit to a model consisting of an exponential plus a constant, and the maximum within each fit window (marked by vertical lines) is extracted. When combined with the background measurement, we obtain the normalized population $P_{\rm S}^{\ket{2}}$ with a typical statistical uncertainty of 0.62\% and 1.9\% for $^{87}$Rb and $^{39}$K, respectively.}
  \label{fig:Detection}
\end{figure}

The normalized output population in $\ket{F = 2}$ is detected for each species by applying a sequence of near-resonant optical pulses. The atomic fluorescence is then collected on an avalanche photodiode (Hamamatsu C12703, $\phi$1.5 mm active area, 10 MHz bandwidth). Figure \ref{fig:Detection} shows an example of a typical detection trace. First, the number of $^{39}$K atoms in $\ket{F=2}$ (labelled $N_{\rm K}^{\ket{2}}$) are detected via a 50 $\mu$s pulse resonant with $\ket{F = 2} \to \ket{F' = 3}$. This signal decays rapidly due to optical pumping into $\ket{F = 1}$ from the nearby $\ket{F'=2}$ state, which makes it necessary to use a high-sensitivity, high-bandwidth detector for $^{39}$K. We fit this signal to an exponential function and extract the peak value as $N_{\rm K}^{\ket{2}}$, as shown in \Fig \ref{fig:Detection}. This is followed by a pulse of the same duration containing optical frequencies resonant with both $\ket{F = 2} \to \ket{F'=3}$ and $\ket{F = 1} \to \ket{F' = 2}$. The intensity ratio of these frequencies is adjusted to obtain as flat a signal as possible---representing an equilibrium in optical pumping rates between the two ground states. This signal is also fit to an exponential function and steady-state value is extracted as $N_{\rm K}^{\rm total}$. An identical protocol is then applied for $^{87}$Rb to obtain $N_{\rm Rb}^{\ket{2}}$ and $N_{\rm Rb}^{\rm total}$. Finally, the normalized population in $\ket{F = 2}$ for each species is determined from
\be
  P_{\rm S}^{\ket{2}} = \frac{N_{\rm S}^{\ket{2}} - B_{\rm S}^{\ket{2}}}{N_{\rm S}^{\rm total} - B_{\rm S}^{\rm total}},
\ee
where $B_{\rm S}^{\ket{2}}$ and $B_{\rm S}^{\rm total}$ are background signals determined from a separate measurement sequence. This background sequence involves the same measurement protocol, except the MOT gradient coils are kept off such that the signal is dominated by stray light and fluorescence from background vapor. We perform one background shot for each set of 82 interferometer shots.

\subsection{Measurement principle and data acquisition protocol}
\label{sec:Protocol}

The primary goal of this experiment is to measure the E\"{o}tv\"{o}s parameter
\be
  \label{eta2}
  \eta = \frac{a_{\rm K} - a_{\rm Rb}}{g},
\ee
where $a_{\rm S}$ is the gravitational acceleration of species S = Rb, K relative to the reference frame defined by the common retro-reflection mirror, and $g$ is the known local gravitational acceleration. The output of each atom interferometer is a sinusoidal fringe pattern given by
\be
  \label{SensorOutput}
  P_{\rm S}^{\ud} = Y_{\rm S}^{\ud} - \frac{C_{\rm S}^{\ud}}{2} \cos \phi_{\rm S}^{\ud},
\ee
where $\uparrow$ ($\downarrow$) represents a momentum transfer in the forward (backward) $z$-direction. The fringe parameters $Y_{\rm S}$ and $C_{\rm S}$ are the offset and contrast, and the phase shift of each interferometer is given by
\be
  \label{phiSUD}
  \phi_{\rm S}^{\ud} = \pm (k_{\rm S} a_{\rm S} - \alpha_{\rm S}) T_{\rm eff}^2 \pm \phi_{\rm S}^{\rm vib} + \phi_{\rm S}^{\ud, \rm sys}.
\ee
 Here, the first term contains the inertial phase of interest $k_{\rm S} a_{\rm S} T_{\rm eff}^2$ and a precise control phase $\alpha_{\rm S} T_{\rm eff}^2$ from the Raman laser frequency chirp. The second term in \Eq \refeq{phiSUD}, $\phi_{\rm S}^{\rm vib}$, is a deterministic phase shift arising from vibrations of the reference mirror. This phase is proportional to the wavevector $k_{\rm S}$, and hence switches sign with opposite momentum transfer. The third term $\phi_{\rm S}^{\ud, \rm sys}$ contains all systematic effects, which we describe in detail in \Sec \ref{sec:Systematics}.

The effective interrogation time $T_{\rm eff}$ is identical for both species. For square Raman pulses with effective Rabi frequency $\Omega_{\rm eff}$ and $\pi/2$-pulse duration $\tau$, $T_{\rm eff}$ can shown to be \cite{Cheinet2008, Merlet2009}
\be
  \label{Teff2}
  T_{\rm eff}^2 = \int f(t) \dd t = (T + 2\tau) \left[ T + \frac{2}{\Omega_{\rm eff}} \tan\left( \frac{\Omega_{\rm eff} \tau}{2} \right) \right],
\ee
where $f(t)$ is atom interferometer response function. In our case, with $T = 20$ ms, $\tau = 2.5$ $\mu$s, and $\Omega_{\rm eff} \tau = \pi/2$, we obtain $T_{\rm eff} = 20.00409$ ms. With identical timing, the ratio of interferometer scale factors is simply
\be
  \label{kappa}
  \kappa \equiv \frac{k_{\rm K}}{k_{\rm Rb}} \simeq 1.017657.
\ee
As mentioned in the previous section, we simultaneously scan the phase of the $^{39}$K and $^{87}$Rb interferometers by independently varying the chirp rates $\alpha_{\rm K}$ and $\alpha_{\rm Rb}$.  We operate the dual-species interferometer by interleaving measurements with $+k_{\rm S}$ and $-k_{\rm S}$ every other shot. This $k$-reversal technique is a well-established method for rejecting non-inertial systematic effects by more than a factor of 100 \cite{LeGouet2008}.

\begin{figure}[!tb]
  \centering
  \includegraphics[width=0.48\textwidth]{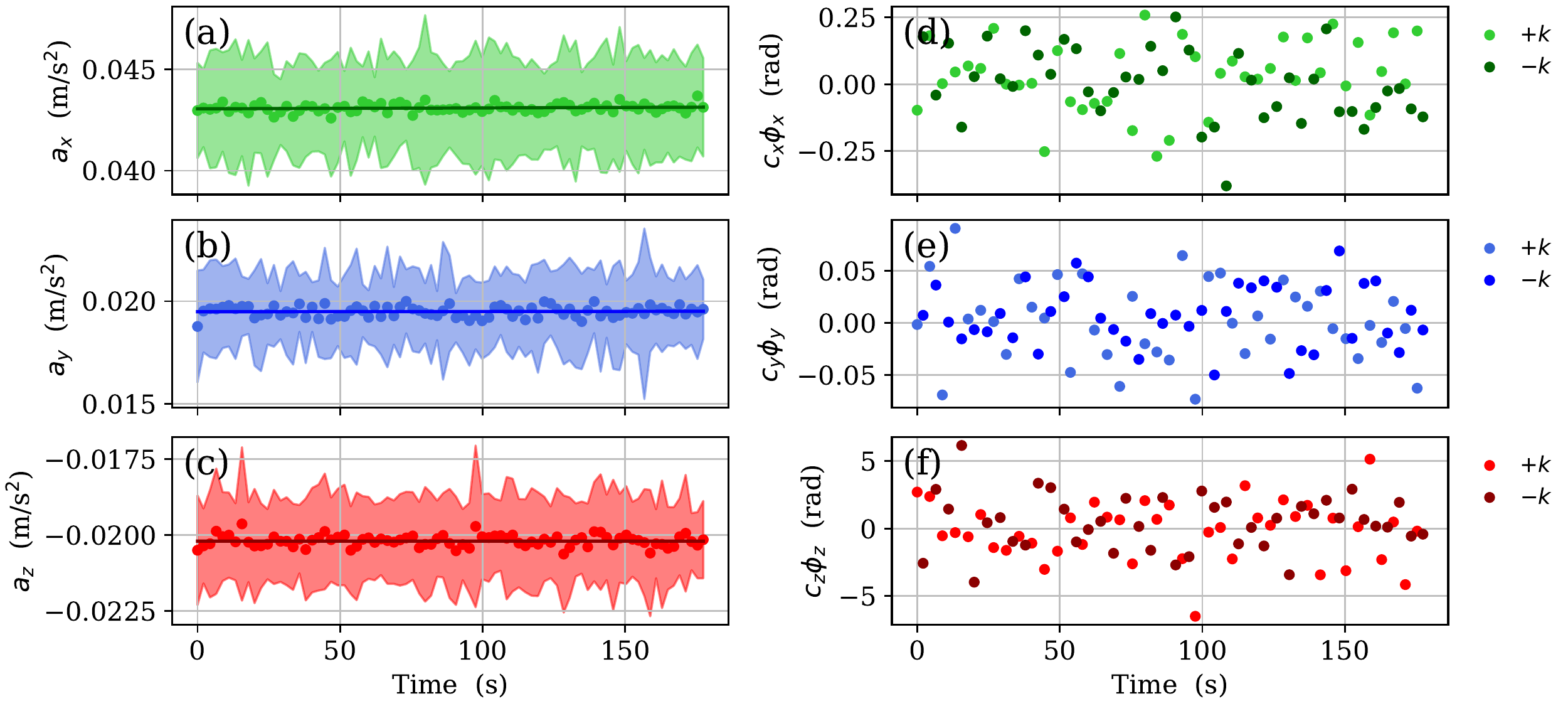}
  \caption{(a--c) Example output from the three-axis MA during one data set of 82 shots. The points represent the mean acceleration over each $2T = 40$ ms interferometer interval, and the shaded area represents the $1\sigma$ standard deviation. Solid lines are linear fits which are used to correct the accelerometer data for a constant bias and a linear drift. Here, the constant bias for the $z$-axis is $b_z \simeq -2.0193(35) \times 10^{-2}$ m/s$^2$. (d--f) Vibration phase estimates for the Rb interferometer corresponding to each accelerometer axis. These phases are multiplied by coefficients $c_x = -0.065$, $c_y = 0.025$, and $1$ ($c_z = 0$) to account for imperfections in the scale factors and alignment of each MA axis.}
  \label{fig:AccelPhases}
\end{figure}

Finally, during each shot, we record the output of a three-axis mechanical accelerometer (MA) (Nanometrics Titan, sensitivity 4.1 V/m/s$^2$, bandwidth 200 Hz) attached to the rear of the retro-reflection mirror. These data allow us to post-correct the vibrational motion of the reference frame using the dual-species fringe reconstruction by accelerometer correlation (FRAC) method we described in previous work \cite{Barrett2015}. Figure \ref{fig:AccelPhases} shows the mean accelerator output and vibration phase estimates for the Rb interferometer. We make a first-order correction of the accelerometer bias drift by subtracting linear fits from the mean output of each axis.\footnote{We note that this correction has a statistical uncertainty of $\delta b \sim 3.6~\mu g$---limited by our knowledge of the mean acceleration at any time. This error produces artificial noise in the phase offset of each interferometer that changes $a_{\rm S}$ randomly between data sets. However, this noise is common to the two interferometers and cancels in the measurement of $\eta$.} The resulting accelerometer signals are then processed as described in Appendix \ref{sec:FRACSignal}, and the vibration phase estimate for each species is computed as
\be
  \label{phivib}
  \tilde{\phi}_{\rm S}^{\rm vib} = k_{\rm S} \int f(t) \big[ \bm{c} \cdot \bm{a}_{\rm vib}(t) \big] \dd t = c_x \phi_{{\rm S}, x}^{\rm vib} + c_y \phi_{{\rm S}, y}^{\rm vib} + (1+c_z) \phi_{{\rm S}, z}^{\rm vib}.
\ee
Here, $\bm{a}_{\rm vib}$ is the AC part of three-axis MA output, and the vector of coefficients $\bm{c} = (c_x, c_y, 1+c_z)$ accounts for imperfections in the scale factor and alignment of each MA axis with respect to the Raman wavevector ($|c_i| \ll 1$). We determine them by simultaneously optimizing the signal-to-noise ratio of all reconstructed fringes (\ie $\pm k_{\rm S}$ for each species). The phase estimate for each species is subtracted from the corresponding interferometer phase on each shot of the experiment. With this method, we are able to reduce the phase noise on each fringe at $T = 20$ ms from $\sigma_\phi \simeq 2.2$ rad (in the absence of a vibration isolation platform) down to $\sim 10$ mrad---corresponding to a vibration rejection ratio $> 200$.

\section{Analysis methods for correlated atom interferometers}
\label{sec:Methods}


In previous work \cite{Barrett2015}, we experimentally demonstrated two methods for extracting the differential phase between dual-species interferometers for precise tests of the UFF. The first method utilizes a MA to reconstruct single-sensor interference fringes based on measurements of the vibration-induced phase. A gravity measurement for each atom interferometer is then performed simultaneously, and the differential acceleration is computed. The second method is based on Bayesian inference, which uses a statistical model and knowledge of the system noise to directly estimate the differential inertial phase between the interferometers. We now describe our analysis results with these two methods.

\subsection{Differential FRAC method}
\label{sec:FRAC}

As mentioned in \Sec \ref{sec:Protocol}, our measurement consists of scanning the frequency chirp $\alpha_{\rm S}$ for each species and applying a vibration phase correction
\be
  \label{tildealpha}
  \tilde{\alpha}_{\rm S} = \alpha_{\rm S} + \tilde{\phi}_{\rm S}^{\rm vib}/T_{\rm eff}^2,
\ee
where $\tilde{\phi}_{\rm S}^{\rm vib}$ is an estimate of the true vibration phase given by \Eq \refeq{phivib}. This allows us to reconstruct the interference fringes on a shot-to-shot basis by plotting the output of each interferometer as a function of $\tilde{\alpha}_{\rm S}$. We independently determine the central fringe by varying $T$ and locating the common dark fringe. We then fit the fringe for each momentum transfer direction, and we extract the chirp rates corresponding to the central fringe:
\be
  \alpha_{\rm S}^{\ud} = \pm k_{\rm S} a_{\rm S} + \phi_{\rm S}^{\ud, \rm sys}/T_{\rm eff}^2.
\ee
To eliminate systematic effects that are insensitive to the direction of momentum transfer, we compute the E\"{o}tv\"{o}s parameter from the half-difference of these central chirp rates:
\be
  \label{etaFRAC}
  \eta^{\rm raw} = \eta + \eta^{\rm sys} = \frac{1}{g} \left[ \left( \frac{\alpha_{\rm K}^{\uparrow} - \alpha_{\rm K}^{\downarrow}}{2k_{\rm K}} \right) - \left( \frac{\alpha_{\rm Rb}^{\uparrow} - \alpha_{\rm Rb}^{\downarrow}}{2k_{\rm Rb}} \right) \right],
\ee
where $\eta^{\rm raw}$ is the raw measured value, $\eta$ is the true value defined by \Eq \refeq{eta2}, and $\eta^{\rm sys}$ is the contribution from systematic effects (see \Sec \ref{sec:Systematics}).

\begin{figure}[!tb]
  \centering
  \includegraphics[width=0.48\textwidth]{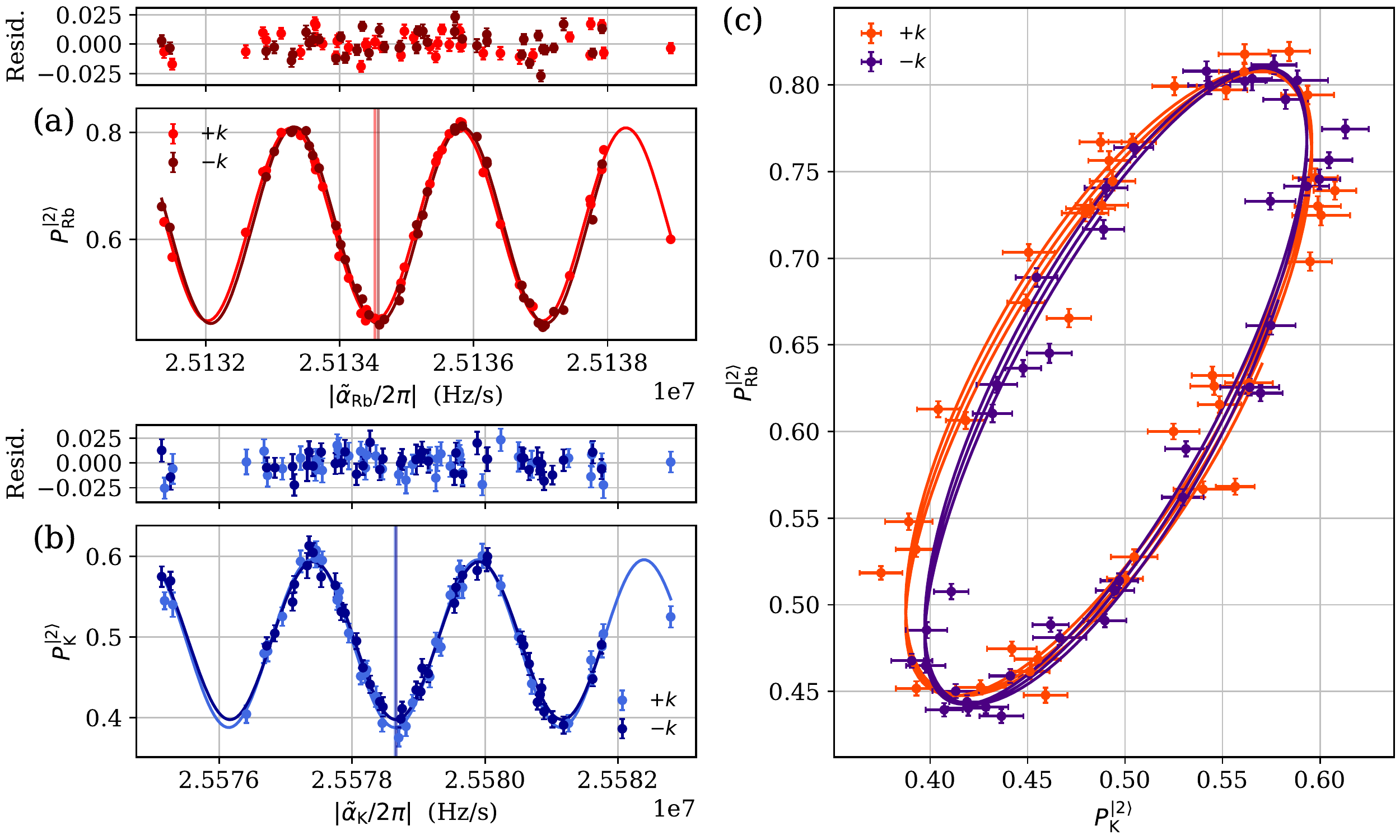}
  \caption{Simultaneous interference fringes for $^{87}$Rb (a) and $^{39}$K (b) as a function of the corrected chirp rate $\tilde{\alpha}_{\rm S}$ at $T = 20$ ms. Two fringes, each containing 41 points, are shown for each species corresponding to opposite momentum transfer directions ($\pm k_{\rm S}$). Data are fit to sinusoidal functions (solid lines), and the fit residuals are shown above each fringe. Error bars correspond to the $1\sigma$ statistical uncertainty obtained from individual detection traces. Vertical lines indicate the central fringe chirp extracted from the fit. (c) Parametric plot of the two interferometer outputs---illustrating the high-level of correlation between species. Curves correspond to the same sinusoidal fits shown in (a) and (b).}
  \label{fig:Fringes}
\end{figure}

Figure \ref{fig:Fringes} shows a typical data set at $T = 20$ ms comprised of 82 shots ($2 \times 41$ for $\pm k_{\rm S}$). Fringes corresponding to each momentum transfer direction are acquired over 172 s using the interleaved protocol described in \Sec \ref{sec:Protocol}. Table \ref{tab:FringeParameters} provides a summary of fringe parameters extracted from these data. We obtain statistical uncertainties of $\delta a_{\rm S}/a_{\rm S} \simeq 1.5 \times 10^{-7}$ ($2.6 \times 10^{-7}$) for each $^{87}$Rb ($^{39}$K) fringe, which results in an uncertainty of $\delta \eta^{\rm raw} \simeq 3.0 \times 10^{-7}$ for each measurement of the E\"{o}tv\"{o}s parameter. These results are limited by the signal-to-noise ratio (SNR) of the reconstructed fringes, which has contributions from the detection noise, phase noise from the laser, and phase noise from the vibration phase estimate\footnote{We note that the rejection of vibration noise using the FRAC method is not a limitation on the differential measurement, but places a large uncertainty on individual acceleration measurements due to an imperfect rejection of the MA bias. This can be solved by filtering out the dc component of the MA.}. In our case, the SNR of the $^{39}$K interferometer is the main limitation on the differential measurement. Yet, despite having 5-times fewer atoms and a lower fringe contrast, the performance of our $^{39}$K interferometer is comparable to $^{87}$Rb.

\begin{table*}[!tb]
  \caption{Summary of fringe parameters extracted from the data shown in \Fig \ref{fig:Fringes}. The SNR is defined as the ratio of the fringe contrast $C$ to the standard deviation of fit residuals. The quantities $\sigma_\Phi$, $\sigma_Y$, and $\sigma_C$ are independent estimates of the phase, offset, and contrast noise, respectively, obtained by minimizing the negative log-likelihood noise model for each interferometer (see Appendix \ref{sec:NoiseEstimation}). The phase noise has contributions from both the laser phase and the FRAC correction due to the MA. We note that $\sigma_C$ is small compared to the other two noise terms. Finally, the gravitational acceleration $a_{\rm S}^{\ud}$ is determined from the central fringe chirp rate $\alpha_{\rm S}^{\ud}$. We emphasize these values are not corrected for systematic effects, and contain a residual error of a few $\mu$g due to the MA bias.}
  \label{tab:FringeParameters}
  \begin{ruledtabular}
  \begin{tabular}{cddddc}
     Quantity       & \multicolumn{1}{c}{$+k_{\rm Rb}$} & \multicolumn{1}{c}{$-k_{\rm Rb}$} & \multicolumn{1}{c}{$+k_{\rm K}$} & \multicolumn{1}{c}{$-k_{\rm K}$} & Unit \\
     \hline
      SNR           & 37.5           & 36.3           & 18.7           & 20.7        & \\
      $C$           & 0.360(5)       & 0.367(5)       & 0.208(6)       & 0.196(5)    & \\
      $\sigma_\Phi$ & 0.0507(13)     & 0.0516(14)     & 0.0564(28)     & 0.0653(30)  & rad \\
      $\sigma_Y$    & 0.0067(1)      & 0.0066(1)      & 0.0085(1)      & 0.0088(1)   & \\
      $\sigma_C$    & 0.0011(1)      & 0.0011(2)      & 0.0048(4)      & 0.0037(4)   & \\
      $\alpha_{\rm S}^{\ud}/2\pi$ & 25.1345323(48) & -25.1345808(54) & 25.5786507(92) & -25.5786696(88) & MHz/s \\
      $a_{\rm S}^{\ud}$ & 9.8055003(19)  & 9.8055192(21)  & 9.8056161(35)  & 9.8056266(34)
      & m/s$^2$ \\
  \end{tabular}
  \end{ruledtabular}
\end{table*}

One advantage of the differential FRAC method is its simplicity---it does not require prior information about the system, such as noise levels and interferometer contrasts, nor does it require an accurate statistical model for the system response. However, it requires a high-sensitivity MA (or seismometer) with good noise characteristics \cite{Merlet2009, Geiger2011, Barrett2015, Menoret2018}. For larger interrogation times, the self-noise of these classical sensors becomes a limitation on the reconstructed phase, and at some point even state-of-the-art devices will not suffice. In this case, an independent method for extracting the differential phase is needed.

\subsection{Generalized Bayesian method}
\label{sec:Bayesian}

Whereas the differential FRAC method can be performed without statistical correlations between atomic sensors, a Bayesian analysis cannot \cite{Stockton2007}. They rely entirely on correlations between the sensor outputs which, when plotted parametrically, form a Lissajous curve as shown in \Fig \ref{fig:Fringes}. When the ratio of scale factors $\kappa = 1$, as it is for gravity gradiometers \cite{Snadden1998, Rosi2014, Biedermann2015, Caldani2019}, the Lissajous curve is an ellipse with an eccentricity determined by the differential phase $\phi_{\rm d}$ between the two sensors. In this simple case, one can extract the phase by using elipse-fitting techniques \cite{Foster2002, Rosi2014}, or by using a Bayesian estimation algorithm \cite{Stockton2007}. The advantage of Bayesian methods is that they are intrinsically optimal and unbiased estimators, whereas ellipse fitting in two dimensions exhibits undesired measurement bias when the differential phase differs from $\pi/2$. In previous work \cite{Barrett2015}, we developed a generalized Bayesian estimator of $\phi_{\rm d}$ for any value of $\kappa$. We briefly describe the principle of this technique in Appendix \ref{sec:BayesianEstimation}.

Since both $Y_{\rm S}$ and $C_{\rm S}$ can be measured independently from the output of each atomic sensor, we begin by rescaling \Eq \refeq{SensorOutput} as $n_{\rm S}^{\ud} = 2(P_{\rm S}^{\ud} - Y_{\rm S}^{\ud})/C_{\rm S}^{\ud}$:
\begin{subequations}
\label{Lissajous}
\begin{align}
  n_{\rm K}^{\ud} & = \cos(\phi_{\rm K}^{\ud}) = \cos(\kappa \phi_{\rm c}^{\ud} + \phi_{\rm d}^{\ud}), \\
  n_{\rm Rb}^{\ud} & = \cos(\phi_{\rm Rb}^{\ud}) = \cos(\phi_{\rm c}^{\ud}).
\end{align}
\end{subequations}
This ensures that the output lies within the unit square and reduces the number of free parameters to estimate with the Bayesian algorithm. We choose $\phi_{\rm c}^{\ud} = \phi_{\rm Rb}^{\ud}$ to be the common phase between the two sensors. The differential phase for each momentum transfer direction is $\phi_{\rm d}^{\ud} = \phi_{\rm K}^{\ud} - \kappa\phi_{\rm Rb}^{\ud}$, where $\kappa$ is the scale factor ratio given by \Eq \refeq{kappa}. From the full phase of each atomic sensor [\Eq \refeq{phiSUD}], the differential phase can be shown to be
\be
  \phi_{\rm d}^{\ud} = \pm k_{\rm K} (a_{\rm K} - a_{\rm Rb}) T_{\rm eff}^2 \mp (\alpha_{\rm K} - \kappa \alpha_{\rm Rb}) T_{\rm eff}^2 + \phi_{\rm d}^{\ud, \rm sys}.
\ee
We emphasize that $\phi_{\rm K}^{\rm vib} = \kappa\phi_{\rm Rb}^{\rm vib}$ since we use identical pulse timing for the two interferometers [see \Eq \refeq{phivib}], hence these terms cancel in $\phi_{\rm d}$. The three remaining terms are the differential inertial phase (which contains the signature of a UFF violation), the precisely-known differential laser phase, 
and the differential systematic phase $\phi_{\rm d}^{\ud, \rm sys} \equiv \phi_{\rm K}^{\ud, \rm sys} - \kappa\phi_{\rm Rb}^{\ud, \rm sys}$. The latter can be partially suppressed by computing the half-difference $\phi_{\rm d} = \frac{1}{2}(\phi_{\rm d}^{\uparrow} - \phi_{\rm d}^{\downarrow})$, which removes direction-independent systematics. The E\"{o}tv\"{o}s parameter is then computed from
\be
  \label{etaBayes}
  \eta^{\rm raw} = \eta + \eta^{\rm sys} = \frac{\phi_{\rm d}}{k_{\rm K} g T_{\rm eff}^2}.
\ee
Both expressions \eqref{etaFRAC} and \eqref{etaBayes} for the E\"{o}tv\"{o}s parameter are equivalent.

To measure $\phi_{\rm d}^{\ud}$, we marginalize the likelihood distribution\footnote{For a derivation of the likelihood distribution in this case, we refer the reader to \Refs \citenum{Barrett2015} and \citenum{Stockton2007}.} over $n_{\rm K}^{\ud}$ and $n_{\rm Rb}^{\ud}$, which both implicitly depend on the common phase $\phi_{\rm c}$. Due to the non-unity scale factor ratio, the algorithm requires as input the approximate range of $\phi_{\rm c}$ in order to estimate $\phi_{\rm d}$ \cite{Barrett2015}. We estimate this phase range based on the sum of the laser phase ($\alpha_{\rm Rb}^{\ud} T_{\rm eff}^2$) and the vibration phase ($\tilde{\phi}_{\rm Rb}^{\rm vib}$) estimated from the MA. The algorithm also requires accurate estimates of the noise parameters for each interferometer. We describe our method for estimating these parameters in Appendix \ref{sec:NoiseEstimation}.

\begin{figure}[!tb]
  \centering
  \includegraphics[width=0.48\textwidth]{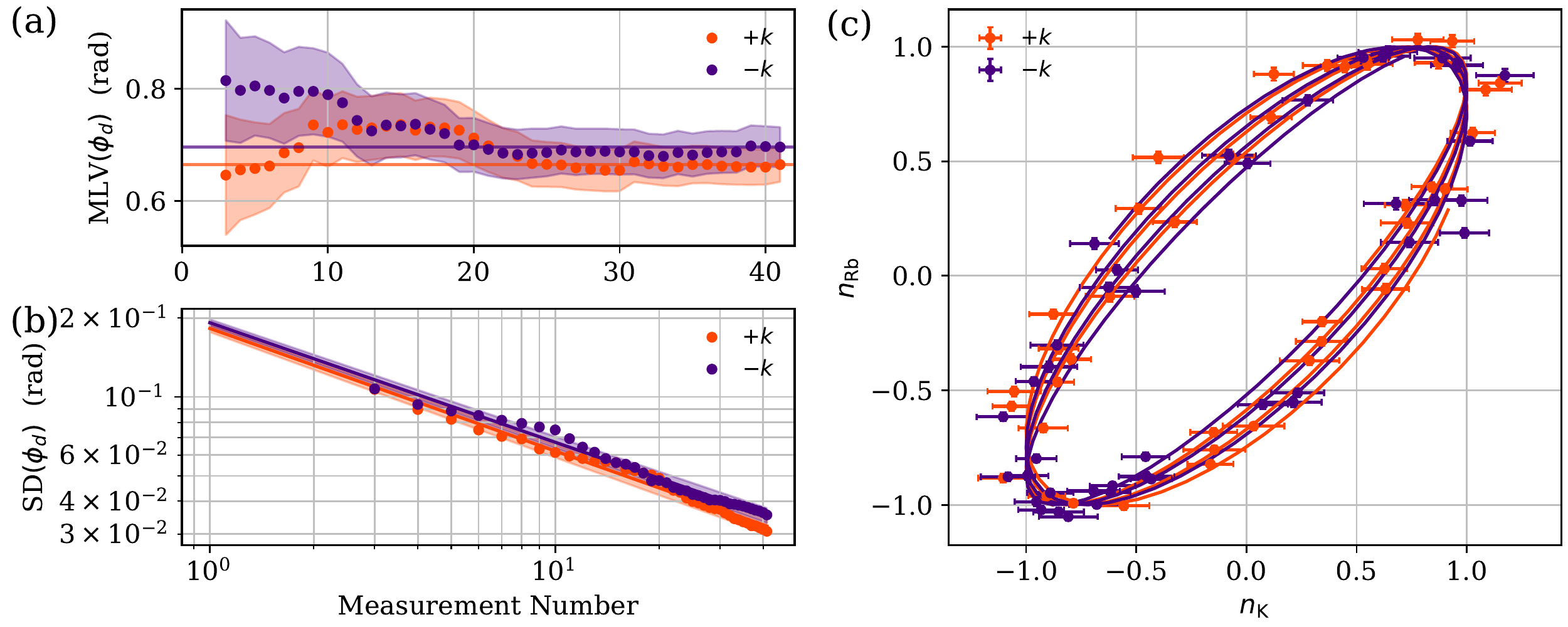}
  \caption{Bayesian analysis results for a single data set consisting of 41 measurements for each momentum transfer direction. Maximum-likelihood value (a) and standard deviation (b) of the differential phase estimate as a function of the measurement number. Shaded regions correspond to $1\sigma$ confidence intervals. Solid lines in (a) indicate the final estimates of $\phi_{\rm d}$. Solid lines in (b) indicate least-squares fits of the form $\Lambda x^{m}$, where we find $\Lambda^{\uparrow} = 0.183(6)$ rad, $m^{\uparrow} = -0.47(1)$, and $\Lambda^{\downarrow} = 0.192(6)$ rad, $m^{\downarrow} = -0.46(1)$. (c) Parametric plot of the scaled interferometer output. Solid curves are Lissajous figures resulting from the final estimates $\phi_{\rm d}^{\uparrow} = 0.665(31)$ rad and $\phi_{\rm d}^{\downarrow} = 0.696(35)$ rad. Input parameters: $\sigma_{\phi_{\rm d}} = 0.053$ rad; $\sigma_{n_{\rm K}} = 0.110$; $\sigma_{n_{\rm Rb}} = 0.022$; approximate range of common phase $\phi_{\rm c} \in [-12, +12]$ rad.}
  \label{fig:BayesianResults}
\end{figure}

Figure \ref{fig:BayesianResults} shows the Bayesian analysis results from a typical data set. As a function of the measurement number, one can clearly observe a rapid convergence of the Bayesian estimation of $\phi_{\rm d}$. As each new measurement is added, the uncertainty in each estimate integrates as $1/\sqrt{N}$---illustrating the optimal nature of the algorithm. The resulting Lissajous figure shows excellent agreement with the scaled sensor output, as shown in \Fig \ref{fig:BayesianResults}(c). We achieve a single-measurement uncertainty $\delta \phi_{\rm d} \simeq 33$ mrad for each momentum transfer direction---corresponding to $\delta\eta^{\rm raw} \simeq 3.7 \times 10^{-7}$ for an interrogation time of $T = 20$ ms. These results compare favorably with those obtained using the differential FRAC method.

\subsection{Correlations and long-term stability}
\label{sec:Correlations}

In this section, we discuss and quantify different types of correlation present in the data. We also present a time-series analysis which reveals the long-term stability and statistical uncertainty of our measurement. For these studies, we acquired a large sample of data over 13.5 h, consisting of 263 individual measurements of $a_{S}^{\ud}$.

\begin{figure}[!tb]
  \centering
  \includegraphics[width=0.48\textwidth]{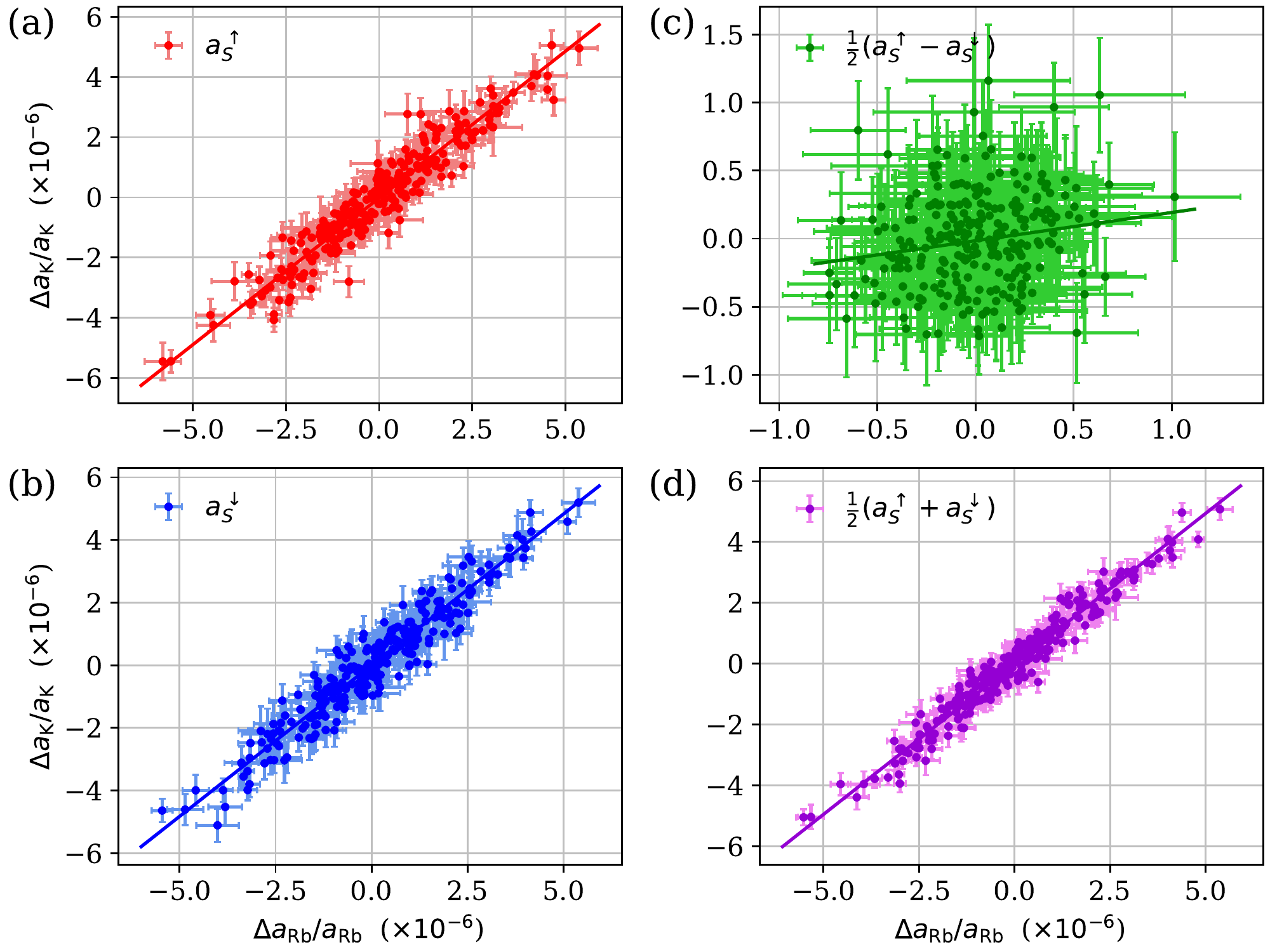}
  \caption{Correlations between acceleration measurements for each species resulting from the differential FRAC method. On each axis we plot the relative change in acceleration $\Delta a/\bar{a} \equiv (a - \bar{a})/\bar{a}$, where $\bar{a}$ is the sample mean. (a,b) Correlations for each momentum transfer direction. (c,d) Correlations for the $k$-independent and $k$-dependent linear combinations of $a_{\rm S}^{\ud}$. Solid lines are linear fits to the data. We obtain correlation coefficients of $R_{\rm Rb, K} = 0.954$ and $0.958$ for (a) and (b), and $R_{\rm Rb,K} = 0.203$ and $0.979$ for (c) and (d), respectively.}
  \label{fig:FRAC-Correlations}
\end{figure}

The differential FRAC method \emph{intrinsically} correlates the two atomic sensors as a consequence of using a single source of information to reconstruct the two interference fringes (\ie the MA). Yet, even with this technique, optimal correlation is still realized when making \emph{simultaneous} measurements due to an efficient common-mode rejection of the time-varying MA bias and various systematic effects. Figure \ref{fig:FRAC-Correlations} shows the correlations between measurements of $a_{\rm Rb}^{\ud}$ and $a_{\rm K}^{\ud}$ resulting from the FRAC method. We quantity the level of correlation with Pearson's correlation coefficient $R_{\rm Rb, K} = \mbox{cov}(a_{\rm Rb}, a_{\rm K})/\sigma_{a_{\rm Rb}} \sigma_{a_{\rm K}}$, where $\mbox{cov}(a_{\rm Rb}, a_{\rm K})$ is the covariance between acceleration data sets $\{a_{\rm Rb}\}$ and $\{a_{\rm K}\}$, and $\sigma_{a_{\rm S}}$ is the corresponding standard deviation of each set. The correlation coefficient can also be interpreted geometrically as the cosine of the angle between two $N$-dimensional vectors represented by these two data sets. Hence, $R_{\rm Rb, K} = \pm 1$ indicates perfect correlation (anti-correlation) when the data are co-linear, while uncorrelated (orthogonal) data produces $R_{\rm Rb, K} = 0$. For the data shown in \Fig \ref{fig:FRAC-Correlations}, we find $R_{\rm Rb, K} > 0.95$ for each momentum transfer direction. This correlation is slightly improved ($R_{\rm Rb, K} \simeq 0.98$) for the $k$-dependent combination of accelerations $\frac{1}{2}(a_{\rm S}^{\uparrow} + a_{\rm S}^{\downarrow})$, due to the rejection of systematic effects that are insensitive to the direction of momentum transfer. Conversely, the $k$-independent combination $\frac{1}{2}(a_{\rm S}^{\uparrow} - a_{\rm S}^{\downarrow})$, which is insensitive to inertial effects, shows very little correlation between species ($R_{\rm Rb, K} \simeq 0.20$). This is expected, as this quantity is comprised primarily of species-specific systematic effects.

The large degree of correlation in acceleration measurements is strongly linked to the bias of the MA, $b$. Our knowledge of this slowly-varying quantity is limited during each measurement ($\delta b \sim 3.6$ $\mu$g, see \Fig \ref{fig:AccelPhases}), which results in a proportional noise on $a_{\rm S}^{\ud}$. However, since this noise is correlated between $^{39}$K and $^{87}$Rb it tends to spread the points along a line with unity slope, as shown in \Fig \ref{fig:FRAC-Correlations}. Since Pearson's correlation coefficient is a measure of the co-linearity between data sets, $|R|$ will tend to one the more extended the data are along this line. Although the accelerometer bias corrupts individual measurements of $a_{\rm S}^{\ud}$, this strong correlation reduces the uncertainty of $\eta$ by almost an order of magnitude, as we show below.

\begin{figure}[!tb]
  \centering
  \includegraphics[width=0.48\textwidth]{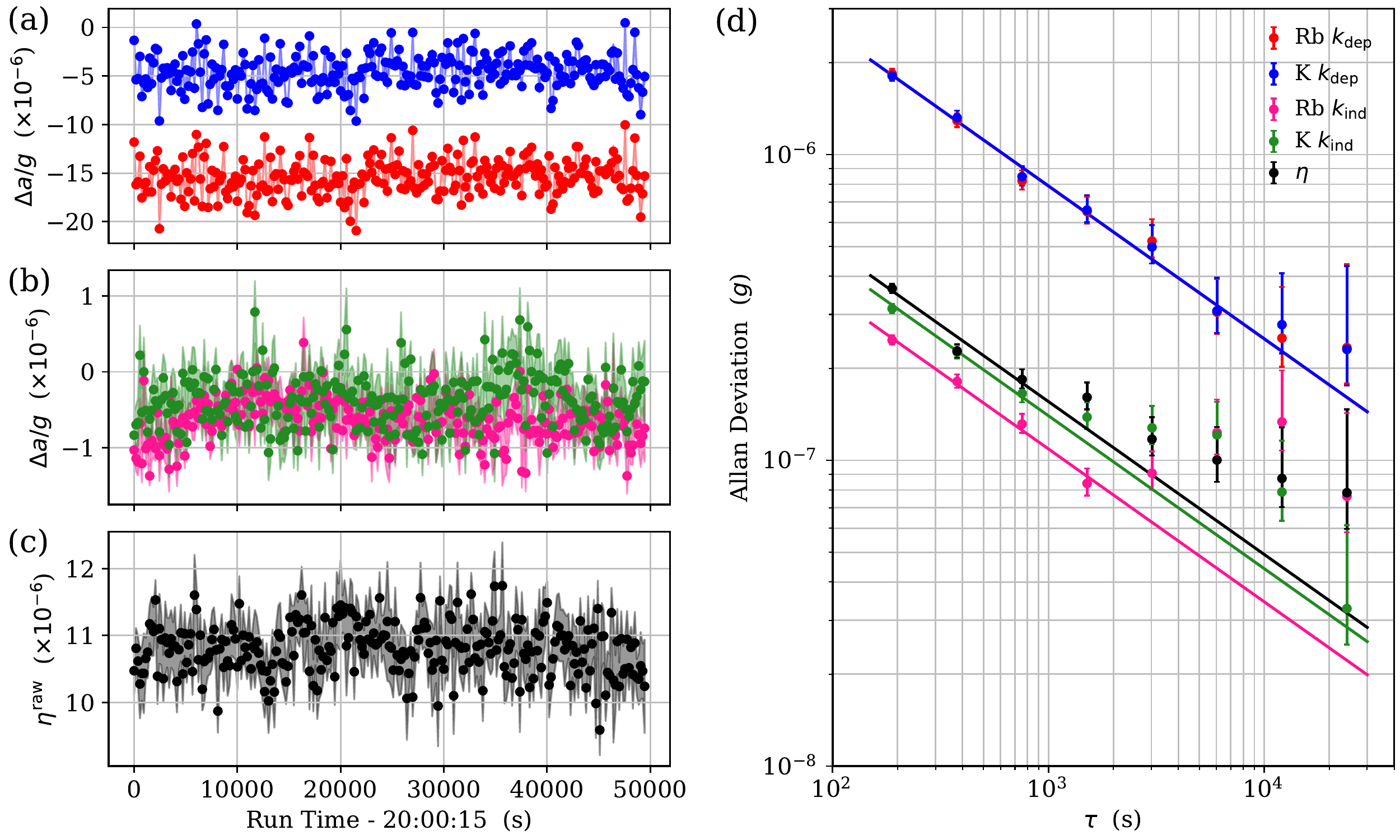}
  \caption{Time-series analysis of UFF measurements obtained with the differential FRAC method. All data are scaled relative to the local value of $g \simeq 9.805642$ m/s$^2$. (a) $k$-dependent accelerations for each species: $(\alpha_{\rm S}^{\uparrow} - \alpha_{\rm S}^{\downarrow})/2k_{\rm S} - g$. (b) $k$-independent accelerations for each species: $(\alpha_{\rm S}^{\uparrow} + \alpha_{\rm S}^{\downarrow})/2k_{\rm S}$. (c) E\"{o}tv\"{o}s parameter deduced from $k$-dependent accelerations [\Eq \eqref{etaFRAC}]. (d) Total Allan deviation of data shown in (a--c). Error bars are computed based on a $\chi^2$-distribution. Solid lines are fits of the form $A/\sqrt{\tau}$. For the $k$-independent accelerations and $\eta^{\rm raw}$, we fit only the first half of the data to illustrate the long-term trend in the absence of experimental drift.}
  \label{fig:FRAC-TimeSeries}
\end{figure}

Figure \ref{fig:FRAC-TimeSeries} shows an analysis of UFF measurements acquired over 13.5 hours. The Allan deviation of $k$-dependent accelerations follow a $1/\sqrt{\tau}$ trend, where $\tau$ is the integration time. However, due to the aforementioned MA bias noise, the standard deviation of these data is large ($\sigma_{a_{\rm S}} \simeq 1.8 \times 10^{-6}~g$) compared to the individual measurement uncertainty $(\sim 2 \times 10^{-7}$). In the presence of measurement correlations, the statistical uncertainty of $\eta$ should be computed as \cite{Barrett2015}
\be
  \label{sigma_eta}
  \sigma_\eta = \frac{1}{g} \left( \sigma_{a_{\rm Rb}}^2 + \sigma_{a_{\rm K}}^2 - 2R_{\rm Rb,K} \sigma_{a_{\rm Rb}} \sigma_{a_{\rm K}} \right)^{1/2}.
\ee
Using $R_{\rm Rb,K} = 0.98$, we find $\sigma_{\eta} = 3.6 \times 10^{-7}$---in perfect agreement with the first point of the Allan deviation of $\eta^{\rm raw}$ shown in \Fig \ref{fig:FRAC-TimeSeries}(d). This Allan deviation decreases as $1/\sqrt{\tau}$ until approximately $10^3$ s. Beyond this time, the trend deviates from the behavior of white frequency noise---showing evidence of a small drift in the measurements. We reach a long-term sensitivity of $\delta\eta^{\rm raw} = 7.8 \times 10^{-8}$ after 24000 s of integration time. We note that measurements of the $k$-independent accelerations [shown in \Figs \ref{fig:FRAC-TimeSeries}(b) and (d)], feature short-term sensitivities at the level of $2.5 \times 10^{-7}~g$ for $^{87}$Rb and $3.1 \times 10^{-7}~g$ for $^{39}$K. These levels are much lower than the corresponding $k$-dependent accelerations due to the cancellation of the MA bias. The Allan deviation of these quantities also shows drift at a similar timescale to $\eta^{\rm raw}$, which is a strong indication that $\eta^{\rm raw}$ is limited by similar drifts of $k$-dependent systematics at long term.

\begin{figure}[!tb]
  \centering
  \includegraphics[width=0.48\textwidth]{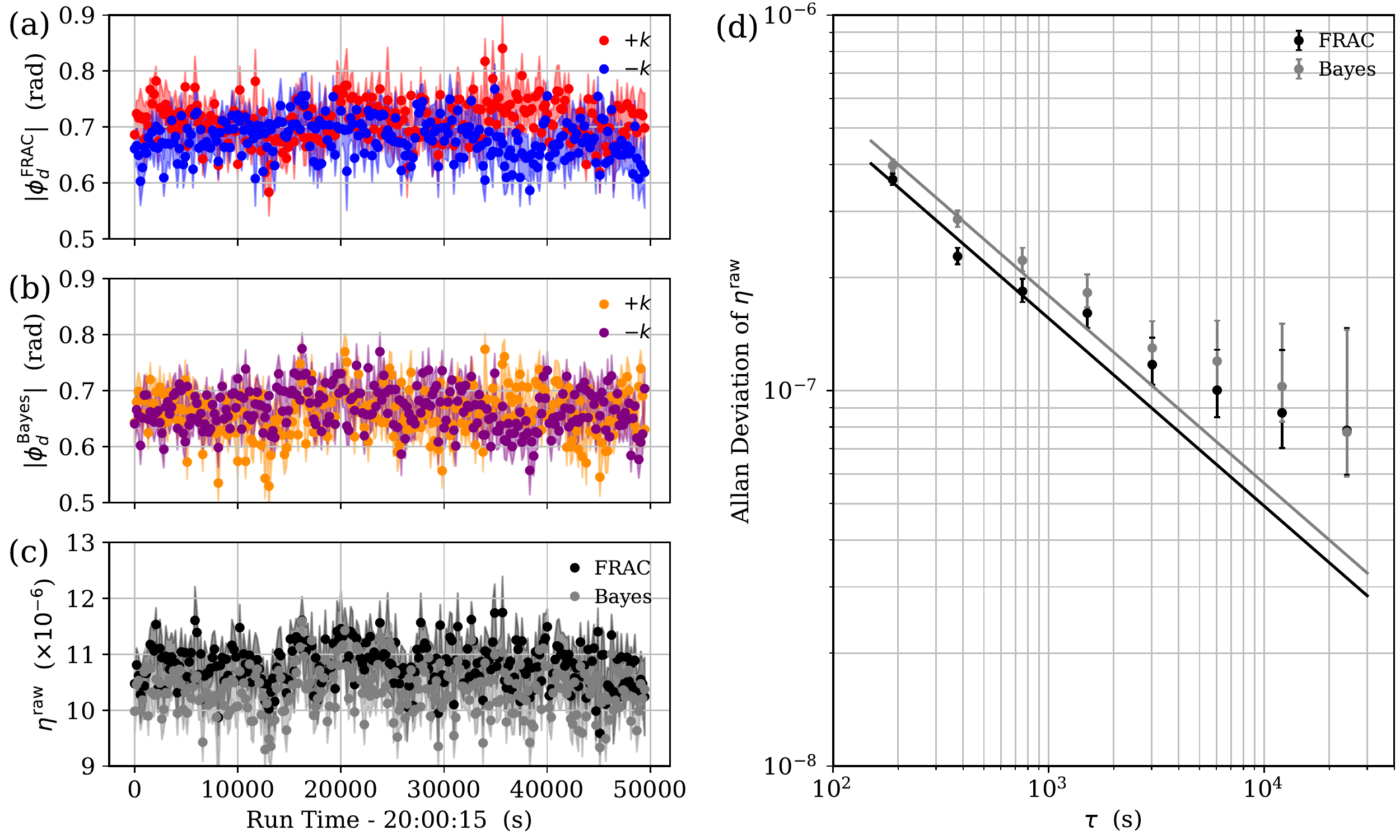}
  \caption{Time-series analysis of Bayesian results, and comparison with FRAC results. (a) Magnitude of the FRAC differential phase for each momentum transfer direction. (b) Magnitude of the Bayesian differential phase for each momentum transfer direction. (c) E\"{o}tv\"{o}s parameter deduced from each method [\Eqs \eqref{etaFRAC} and \eqref{etaBayes}]. (d) Total Allan deviation of $\eta^{\rm raw}$ for both methods. Error bars are computed based on a $\chi^2$-distribution. Solid lines are fits of the form $A/\sqrt{\tau}$ to the first half of each data set.}
  \label{fig:FRAC-Bayes-TimeSeries}
\end{figure}

We now compare these results to those obtained using the generalized Bayesian analysis on the same data. Figure \ref{fig:FRAC-Bayes-TimeSeries} shows a time-series analysis of the differential phase obtained from the two methods for each momentum transfer direction, along with the deduced E\"{o}tv\"{o}s parameter. The results are remarkably similar---to the extent that $\phi_{\rm d}^{\ud, \rm Bayes}$ even reproduces the small-scale features present in $\phi_{\rm d}^{\ud, \rm FRAC}$. Here, the differential phase from the FRAC method is computed based on the measured accelerations:
\be
  \label{phidFRAC}
  \phi_{\rm d}^{\ud, \rm FRAC} = \pm k_{\rm K} \left( a_{\rm K}^{\ud} - a_{\rm Rb}^{\ud} \right) T_{\rm eff}^2.
\ee
To more easily compare these two phases, and those from the Bayesian analysis, we plot their magnitude in \Figs \ref{fig:FRAC-Bayes-TimeSeries}(a,b). For both methods, the E\"{o}tv\"{o}s parameter is then deduced from the $k$-dependent combination of differential phases according to \Eq \refeq{etaBayes}. We observe a small offset of 37 mrad between the Bayesian and FRAC results. This may arise due to our limited knowledge of the effective interrogation time $T_{\rm eff}$, which is derived from the sensitivity function of the interferometer. This function assumes the Raman pulses have a square intensity profile with fixed Rabi frequency and ideal durations ($\Omega_{\rm eff} \tau = \pi/2$). In a real experiment, imperfections in the pulse shapes and intensities affect our knowledge of the true $T_{\rm eff}$, and hence the measurement of $\eta$ through $\phi_{\rm d}$. A relative error of only $\delta T_{\rm eff}/T_{\rm eff} = 3 \times 10^{-7}$ is enough to explain the observed offset. One advantage of working in terms of chirp rate (as with most atomic gravimeters\cite{LeGouet2008, Altin2013, Hu2013, Freier2016, Menoret2018}) is that the measurement of acceleration is insensitive to $T_{\rm eff}$. This is because the interferometer is operated around the central fringe where the total phase is zero, and $a_{\rm S}^{\ud} = \alpha_{\rm S}^{\ud}/k_{\rm S}$.

From the Allan deviation in \Fig \ref{fig:FRAC-Bayes-TimeSeries}(c), the Bayesian estimate of $\eta^{\rm raw}$ shows a slight degradation in terms of the short-term sensitivity ($4.0 \times 10^{-7}$ at 200 s) compared to the differential FRAC method ($3.5 \times 10^{-7}$ at 200 s). Yet the measured sensitivities at long term are consistent within their respective uncertainties---indicating that they are both limited by drifts in systematic effects.

The consistency of these results is paramount for future high-sensitivity tests of the UFF. For the first time, we show that with highly-correlated dual-species interferometers the E\"{o}v\"{o}s parameter can be measured with high precision in the absence of phase stability. This has important implications for future long-baseline tests of the UFF \cite{Hartwig2015, Overstreet2018}, where free-fall times $> 1$ s are planned. At these levels, achieving a phase-stable inertial reference frame will be extremely challenging. Similar to ellipse-fitting methods for atomic gradiometers \cite{Foster2002, Rosi2015}, our Bayesian estimation method completely removes this requirement.

\section{Study of systematic effects}
\label{sec:Systematics}


In this section, we provide a detailed description of our systematic study. Many of the systematic effects we discuss are common to other atom interferometers, but we highlight the additional difficulties due to the physical properties of $^{39}$K that lead to some surprising effects.

Following the analysis from \Sec \ref{sec:FRAC}, the contribution to the E\"{o}tv\"{o}s parameter from systematic effects can be shown to be
\be
  \label{etasysFRAC}
  \eta^{\rm sys} = \frac{1}{g} \left[ \left( \frac{\phi_{\rm K}^{\uparrow, \rm sys} - \phi_{\rm K}^{\downarrow, \rm sys}}{2k_{\rm K} T_{\rm eff}^2} \right) - \left( \frac{\phi_{\rm Rb}^{\uparrow, \rm sys} - \phi_{\rm Rb}^{\downarrow, \rm sys}}{2k_{\rm Rb} T_{\rm eff}^2} \right) \right],
\ee
where $\phi_{\rm S}^{\ud,\rm sys}$ is the sum of all systematic phase shifts corresponding to a given species and momentum transfer direction. Equation \refeq{etasysFRAC} can be written equivalently in terms of differential systematic phase shifts as
\be
  \label{etasysBayes}
  \eta^{\rm sys} = \frac{\phi_{\rm d}^{\uparrow, \rm sys} - \kappa\phi_{\rm d}^{\downarrow, \rm sys}}{k_{\rm K} g T_{\rm eff}^2}.
\ee
For each species, many of the systematic effects we discuss below are coupled because they depend on the same physical parameters, such as the atomic velocity, atom-mirror distance, magnetic field, and Rabi frequency. It was therefore necessary to independently determine these parameters, and include this coupling in our model for each systematic.

\subsection{Velocity sensitivity of the detection system}
\label{sec:AtomicVelocity}


The atomic velocity is a critical parameter for most systematic effects. We found that our detection system has a strong influence on the atomic trajectories that contribute to our data. Most detectors implemented on atom interferometers collect atomic fluorescence from an observation zone defined by the geometry of the detection optics and the photodetector---making them spatially selective to a certain degree. By flashing on a short detection pulse at a specific time, our detector's spatial selection converts to a velocity selection. Although this effect is known in atomic gravimeters \cite{Gillot2016}, it has not been well modelled or quantified---particularly for dual-species experiments where the atomic trajectories can be significantly different.

Using a statistical approach similar to \Ref \citenum{Yavin2002}, we have developed an analytical model for the velocity shift produced by our detection system. We provide a brief description of the model below, further details will be published elsewhere. We use a phase-space representation of the atom cloud's probability density with a 3D Maxwell-Boltzmann distribution of velocities and a 3D Gaussian distribution of positions. This distribution is propagated in time while in free-fall under gravity---causing the spatial width to undergo hyperbolic expansion and the center position to follow a parabolic trajectory along the vertical. The detection signal at time $t$ is obtained by integrating over the density of atoms contained within a sphere of radius $r_{\rm d}$ located a distance $z_{\rm d}$ below the initial position of the cloud. This signal is strongly influenced by the ratio of $r_{\rm d}$ to the cloud size at the time of detection, as well as the time at which the detection takes place. To avoid biasing the signal toward a particular velocity class, the signal should be acquired when center of the cloud is located at the center of the detection zone. However, this is not always possible (\eg due to geometric or timing constraints), in which case one should aim to detect as much of the cloud as possible. 

Using this model for the detection signal, we obtained an analytical expression for the velocity shift produced by the detection system in terms of measurable parameters such as the cloud size and temperature. Based on a simple averaging argument, we consider this shift the relevant parameter for several velocity-dependent systematic effects in the atom interferometers. The velocity shift is given by
\be
  \label{Deltavzpm}
  \Delta v_{\rm S}^{\ud}(t)
  = -\frac{1 - \gamma \varsigma e^{-\mu_{\rm f}^2} \mbox{erf}(\mu_{\rm i}/\varsigma)}{1 - \gamma \cdot \mbox{erf}(\mu_{\rm i})} \frac{\Delta z_{\rm S}^{\ud}(t)}{t},
\ee
with the following dimensionless parameters
\be
  \gamma      = \frac{\sqrt{\pi}}{2 \mu_{\rm i}} e^{\mu_{\rm i}^2}, \;\;\;\;\;\;\;
  \varsigma   = \frac{\Sigma}{\sigma_v t}, \;\;\;\;\;\;\;
  \mu_{\rm i} = \frac{r_{\rm d}}{\sigma_r}, \;\;\;\;\;\;\;
  \mu_{\rm f} = \frac{r_{\rm d}}{\Sigma}.
\ee
Here, $\sigma_r$ is the initial cloud radius, $\sigma_v = \sqrt{2 k_{\rm B} \mathbb{T}/M}$ is the velocity width of the cloud at temperature $\mathbb{T}$, and $\Sigma(t) = \sqrt{\sigma_r^2 + (\sigma_v t)^2}$ is width of the cloud at time $t$. Although we have omitted the subscript $\rm S$ for simplicity, all of these cloud parameters are species dependent. Finally, $\Delta z_{\rm S}^{\ud}(t)$ is the shift in atomic position\footnote{Equation \eqref{Deltazpm} assumes the clouds corresponding to states $\ket{F=1}$ and $\ket{F=2}$ are not significantly separated in space during the detection pulse at time $t$.} relative to the center of the detection zone $z_{\rm d}$
\be
  \label{Deltazpm}
  \Delta z_{\rm S}^{\ud}(t) = \frac{1}{2} g t^2 \pm \frac{1}{2} v_{\rm S}^{\rm rec} (t - {\rm TOF}) - z_{\rm d}.
\ee

\begin{figure}[!b]
  \centering
  \subfigure{\includegraphics[width=0.23\textwidth]{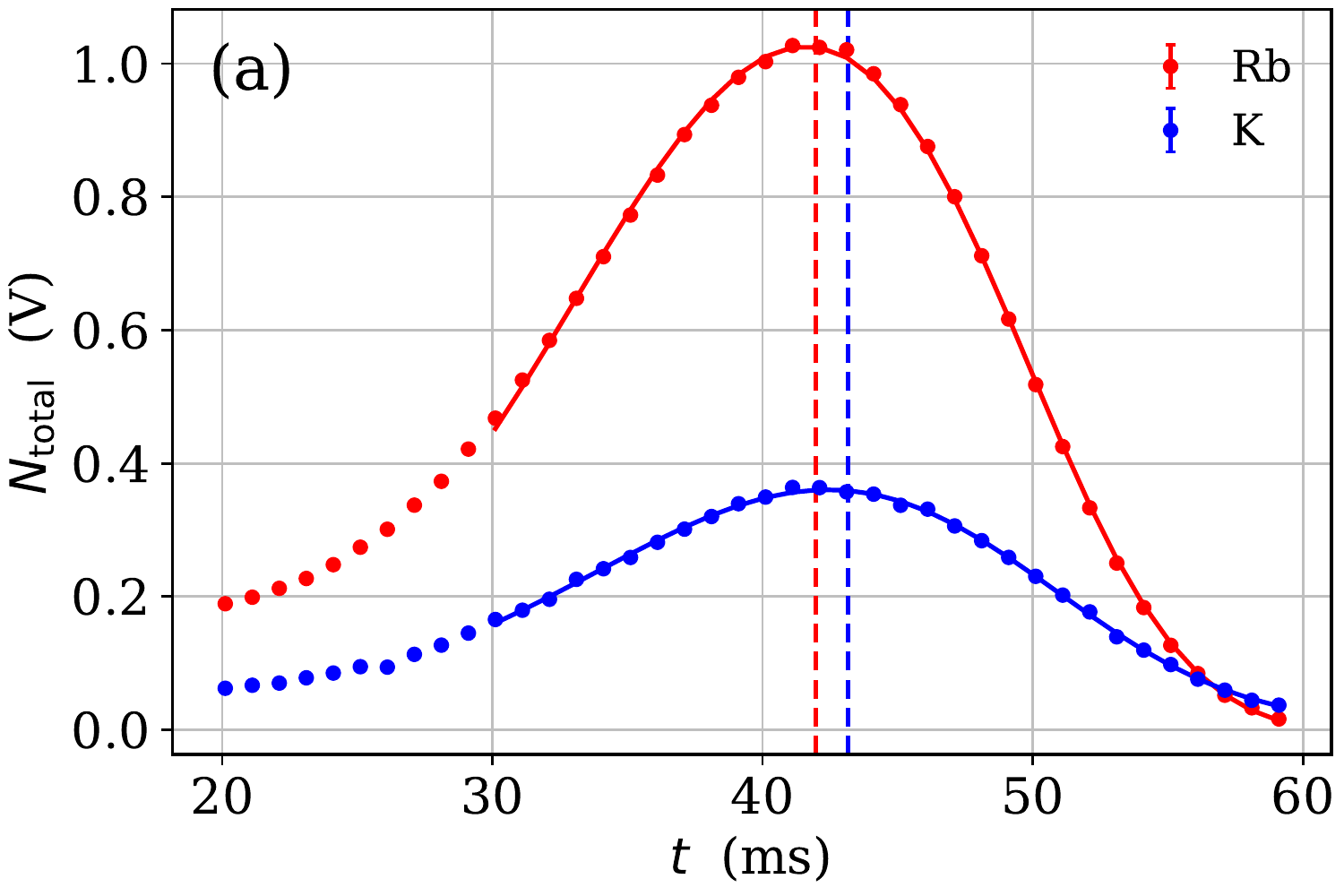}}
  \subfigure{\includegraphics[width=0.23\textwidth]{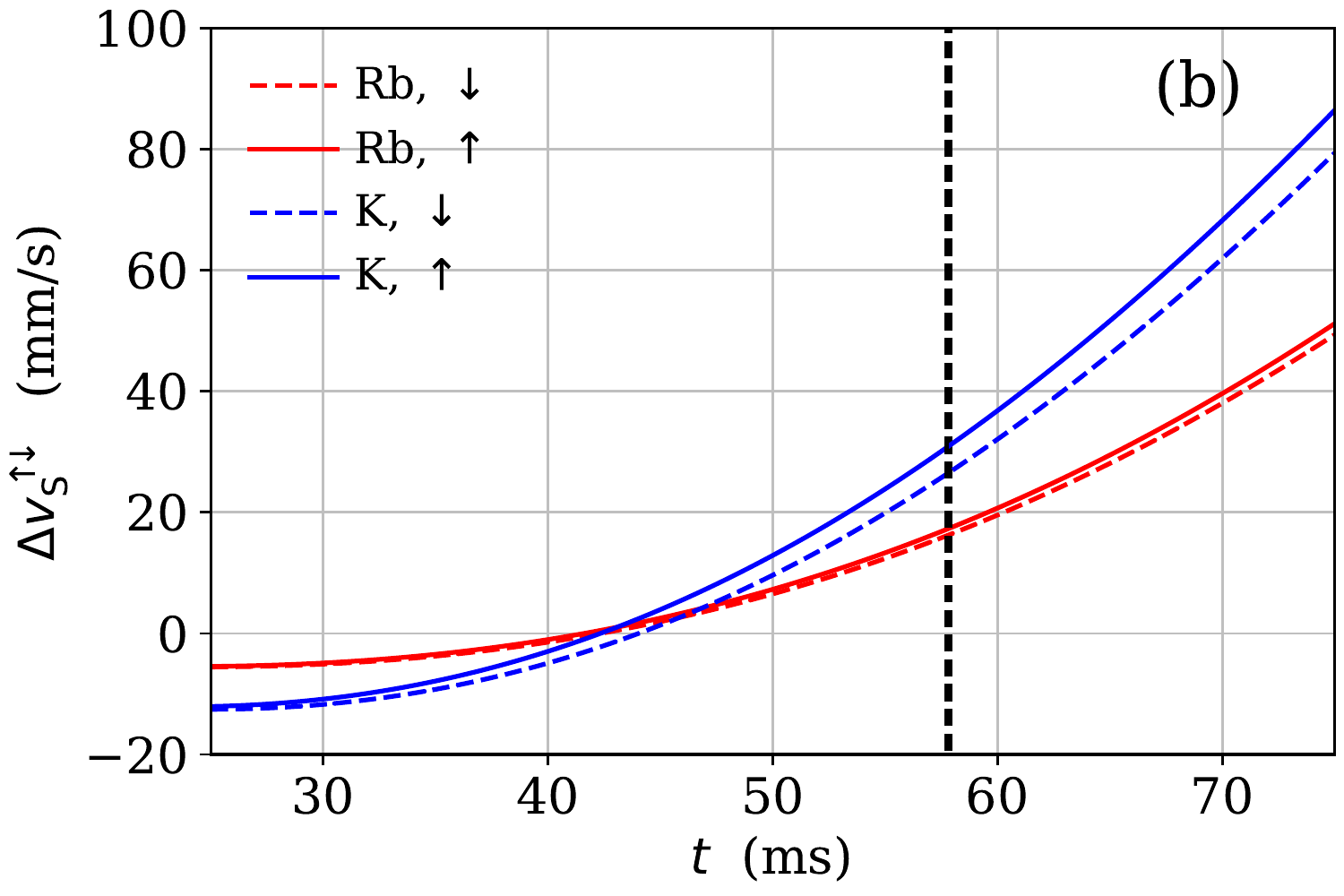}}
  \caption{(a) Measured detection profiles for $^{87}$Rb and $^{39}$K samples. These data represent the total number of atoms detected along the unperturbed trajectory (i.e. no Raman transitions were made before the detection pulse). These curves provide a means to calibrate both the atomic cloud and detection system parameters. The vertical dashed lines indicate the center of the detection profiles obtained from the fits: $t_{\rm d} = \sqrt{2 z_{\rm d}/g}$. (b) Detection-induced velocity shift for each interferometer as a function of the detection time. The vertical dashed line indicates the detection time used in our experiments.}
  \label{fig:MeasuredDetection}
\end{figure}

To evaluate the velocity shift in our case, we require estimates of the parameters $\sigma_r$ and $\sigma_v$ for each species, as well as $r_{\rm d}$ and $z_{\rm d}$ for the detection system. To obtain these parameters, atoms are released after the molasses cooling stage and subsequently imaged by the detection system as a function of the time in free fall. Figure \ref{fig:MeasuredDetection}(a) shows the resulting detection profile for two simultaneous samples of $^{87}$Rb and $^{39}$K at temperatures of $\mathbb{T}_{\rm Rb} = 5.0(5)$ $\mu$K and $\mathbb{T}_{\rm K} = 4.7(5)$ $\mu$K, respectively. These data are fit to our model of the integrated probability density with $t_{\rm d}$, $r_{\rm d}$, and $\sigma_r$ as free parameters. The velocity width for $^{87}$Rb and $^{39}$K was fixed at $\sigma_v = 31.0$ and 44.8 mm/s, respectively, as determined from separate velocity-sensitive Raman spectra. Table \ref{tab:DetectionParameters} lists the best fit parameters for the curves shown in \Fig \ref{fig:MeasuredDetection}(a). We note that this procedure does not require any a priori information about the detection system\footnote{Our detector is an Avalanche Photodiode (APD) Hamamatsu C12703 with a diameter of 1.5 mm. This solution has been chosen because a fast photodiode is required for Potassium. Indeed the detection of the two internal state is done sequentially due to the compactness of the experiment and the signal of $F = 2$ is weak because of the strong depumping rate due to potassium's compact level structure.} or the atomic clouds.

\begin{table}[!tb]
  \caption{Detection profile fit results and the corresponding velocity shift for each atom interferometer. The velocity shift assumes the interferometer sequence starts after a time-of-flight ${\rm TOF} = 16.2$ ms and the detection occurs at $t = {\rm TOF} + 2T + 1.5$ ms = 57.7 ms with $T = 20$ ms.}
  \label{tab:DetectionParameters}
  \begin{ruledtabular}
  \begin{tabular}{cccc}
    Parameter   & $^{87}$Rb & $^{39}$K   & Unit \\ \hline
    $z_{\rm d}$ & 8.634(10) & 9.133(14)  & mm  \\
    $r_{\rm d}$ & 4.288(43) & 4.300(50)  & mm \\
    $\sigma_r$  & 2.83(12)  & 3.32(11)   & mm \\
    $\sigma_v$  &  31.0     &  44.8      & mm/s \\ \hline
    $\Delta v_{\rm S}^{\uparrow}$ & 17.4(6.9) & 30.7(6.0) & mm/s \\
    $\Delta v_{\rm S}^{\downarrow}$ & 16.4(6.5) & 26.3(5.1) & mm/s \\
  \end{tabular}
  \end{ruledtabular}
\end{table}

Using these parameters along with our experimental timing, we estimate a detection-induced velocity shift of $\Delta v_{\rm Rb}^{\ud} \sim 1.5 \, v_{\rm Rb}^{\rm rec}$ for $^{87}$Rb, and $\Delta v_{\rm K}^{\ud} \sim 1.0 \, v_{\rm K}^{\rm rec}$ for $^{39}$K, where $v_{\rm S}^{\rm rec} = \hslash k_{\rm S}/M_{\rm S}$ is the recoil velocity, as listed at the bottom of Table \ref{tab:DetectionParameters}. The difference between the shifts for each momentum transfer direction is due to the different spatial positions of the cloud at the time of detection. Finally, we show the variation in the velocity shift as a function of the detection time $t$ in \Fig \ref{fig:MeasuredDetection}(b). The shift increases quadratically with $t$ according to \Eqs \eqref{Deltavzpm} and \eqref{Deltazpm}, with a zero when the clouds are at the center of the detection zone. The effect is stronger in potassium due to its larger cloud size and velocity spread. 

In the following of this section, these velocity shifts are taken into account in the evaluation of the following systematic errors: second-order Zeeman effect, parasitic lines, two-photon light shift, gravity gradient, scale factor.

\subsection{Second-order Zeeman effect}
\label{sec:QuadraticZeeman}

The interferometers are constructed from ground states with magnetic quantum number $m_F = 0$. Although there is no $1^{\rm st}$-order Zeeman shift, the $2^{\rm nd}$-order Zeeman effect is the most significant contribution to our measurement of $\eta$. This is because the magnitude of this effect is $\sim 15$ times stronger in $^{39}$K compared to $^{87}$Rb and, in the presence of a spatial gradient $\nabla B$, atoms following the two trajectories associated with $\pm k_{\rm S}$ experience slightly different $B$-fields. Hence, the phase shift is not perfectly rejected by reversing the direction of momentum transfer.

\begin{figure}[!tb]
  \centering
  \includegraphics[width=0.48\textwidth]{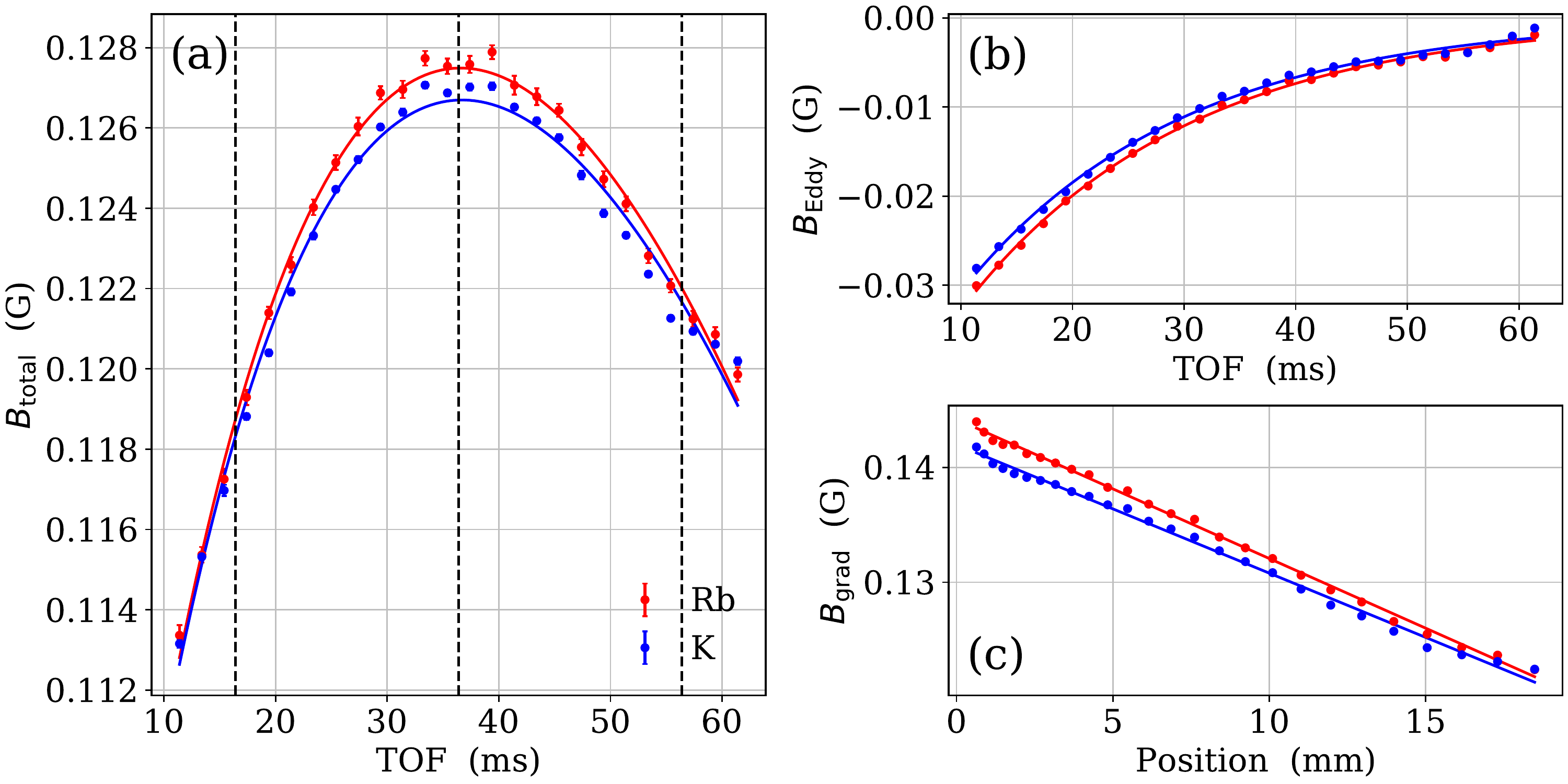}
  \caption{(a) Profile of the magnetic field probed by $^{87}$Rb (red) and $^{39}$K (blue) as a function of the time-of-flight after molasses release. (b) Time-varying component of the field due to Eddy currents. (c) Spatially-varying component of the field due to a gradient. Vertical grid lines on (a) correspond to the interferometer pulses times.}
  \label{fig:MagneticField}
\end{figure}

The $2^{\rm nd}$-order Zeeman shift of the clock transition $\ket{F=1,m_F=0} \to \ket{F=2,m_F=0}$ is $\Delta \omega_{\rm S}^B = 2\pi K_{\rm S} B^2$, where $K_{\rm S}$ is a constant that depends on Land\'{e} $g$-factors and the hyperfine ground-state splitting $\Delta\omega_{\rm S}^{\rm HF}$
\be 
  \label{KS}
  K_{\rm S} = \frac{(g_J - g_I)^2 \mu_B^2}{2\Delta\omega_{\rm S}^{\rm HF}} = \left\{
  \begin{array}{rl}
  575.15 \mbox{ Hz/G}^2 & \mbox{for } ^{87}\rm{Rb}, \\
  8513.75 \mbox{ Hz/G}^2 & \mbox{for } ^{39}\rm{K}. \\
  \end{array} \right.
\ee
We deduce the phase shift on the atom interferometers from
\be
  \label{ZeemanPhase}
  \phi^{\ud,\rm Zeeman}_{\rm S} = 2\pi K_{\rm S} \int g_{\rm S}(t) \big( B_{\rm S}^{\ud}(t) \big)^2 \dd t,
\ee
where $g_{\rm S}(t)$ is the sensitivity function\cite{Cheinet2008} and $B_{\rm S}^{\ud}(t)$ is the time-varying magnetic field experienced by species ${\rm S = Rb, K}$ along the two center-of-mass trajectories associated with opposite momentum transfer. To evaluate this phase shift, the magnetic field profile is measured directly with the atoms by making simultaneous two-photon Raman spectroscopy with both species. We prepare both samples in a mixture of the three internal states $\ket{F=1,m_F=0}$ and $\ket{F=1,m_F=\pm 1}$, and we measure the spectrum of velocity-insensitive $\sigma^+$ Raman transitions along the vertical axis. The frequency splitting between resonances associated with magnetically-sensitive states is directly linked to the magnetic field along the \emph{undiffracted} atomic trajectory during the interferometer sequence. We measure the frequency splitting for each species as a function of the time-of-flight (TOF), and we solve the Breit-Rabi equation\cite{Steck2019} to obtain the magnetic field at each time. The results are shown in \Fig \ref{fig:MagneticField}(a), where the uncertainty of each measurement is $\sim 0.2$ mG. At this level, we note that the influence of the vector and tensor light shifts must be included in order to find agreement between the magnetic field values given by each species\cite{Hu2017}. For our experimental parameters, the sum of vector and tensor light shifts is $-4.5$ kHz for $^{87}$Rb and $-6.7$ kHz for $^{39}$K---corresponding to virtual magnetic fields of $-3.2$ and $-4.8$ mG, respectively.

The magnetic field profile shown in \Fig \ref{fig:MagneticField}(a) consists of a temporally-varying component due to Eddy currents, and a spatially-varying component due to a magnetic field gradient. Eddy currents are induced in a nearby aluminum breadboard when the magnetic bias field and the MOT gradient field are pulsed on or off. We servo-lock the field using a flux-gate sensor near the atoms, which stabilizes the Eddy currents produced by the bias coils $\sim 5$ ms after turning on the bias. However, a residual exponentially-decaying Eddy field persists due to the MOT coils, which we verified separately by keeping the bias coils permanently on during the sequence.

To compute the phase shift of the atom interferometers for opposite momentum transfer, it's important to isolate the temporal and spatial components of the magnetic field. To achieve this, we fit the data shown in \Fig \ref{fig:MagneticField}(a) to the following model:
\be
  \label{BModel}
  B(t) = B_0 + \nabla B_z \left( v_0 t + \frac{1}{2} g t^2 \right) + A_{\rm Eddy} e^{-\Gamma_{\rm Eddy} t},
\ee
where $B_0$ is a constant offset, $\nabla B_z$ is the vertical components of the gradient, $v_0$ is the initial velocity of the sample (which we take to be zero for this analysis), $A_{\rm Eddy}$ is the amplitude of Eddy currents, and $\Gamma_{\rm Eddy}$ is their decay rate. This model assumes that the atoms experience the same Eddy field everywhere in space, which is a good approximation for small displacements relative to the size of the coils. We have separately confirmed that this model accurately represents the field by applying a known gradient and extracting it's value from the fits. Figures \ref{fig:MagneticField}(b) and (c) show the temporal and spatial components, respectively, that are derived from the fits in \Fig \ref{fig:MagneticField}(a). The best fit parameters for each species are consistent within their statistical uncertainties, hence we average them together to obtain the following field parameters: $B_0 = 143.24(78)$ mG, $\nabla B_z = -1.172(33)$ G/m, $A_{\rm Eddy} = -52.8(1.0)$ mG, and $\Gamma_{\rm Eddy} = 50.25(38)$ s$^{-1}$. The field profiles during the interferometer $B_{\rm S}^{\ud}(t)$ are then obtained from \Eq \eqref{BModel} by adding $\pm \frac{1}{2} v_{\rm S}^{\rm rec} (t - \mbox{TOF})$ to the term proportional to $\nabla B_z$. This accounts for the shift in the center-of-mass position due to the photon recoil during the interferometer. Table \ref{tab:Zeeman} summarizes the phase shifts due to the 2nd-order Zeeman effect.

\begin{table}[!tb]
  \caption{Phase shifts due to the 2nd-order Zeeman effect for the following experimental parameters: TOF = 16/2 ms, $T = 20$ ms, $v_{\rm Rb}^{\uparrow} = 1.74(69)$ cm/s, $v_{\rm K}^{\uparrow} = 3.07(60)$ cm/s, $v_{\rm Rb}^{\downarrow} = 1.64(65)$ cm/s, $v_{\rm K}^{\downarrow} = 2.63(51)$ cm/s, $B_0 = 143.24(78)$ mG, $\nabla B_z = -1.172(33)$ G/m, $A_{\rm Eddy} = -52.8(1.0)$ mG, and $\Gamma_{\rm Eddy} = 50.25(38)$ s$^{-1}$.}
  \label{tab:Zeeman}
  \begin{ruledtabular}
  \begin{tabular}{cccccc}
      Species & $\phi_{\rm S}^{\uparrow, \rm Zeeman}$ & $\phi_{\rm S}^{\downarrow, \rm Zeeman}$ & Units \\
    \hline
     $^{87}$Rb & $6.2(6.0)$ & $11.5(5.9)$ & mrad \\
     $^{39}$K  & $-39(91)$  & $151(88)$   & mrad \\
  \end{tabular}
  \end{ruledtabular}
\end{table}

\subsection{Parasitic lines}
\label{sec:ParasiticLines}

In the presence of additional laser frequencies separated by the two-photon resonance, each optical pulse in the atom interferometer diffracts atoms along parasitic trajectories which experience a different phase shift. For sufficiently cold atoms, these trajectories lie within the coherence length of the interference pattern and their phase shifts add a bias to the interferometer \cite{Carraz2012}. This systematic effect is present only in $^{87}$Rb due to the use of an electro-optic phase modulator (PM) to generate the second Raman frequency. At a value of approximately $-0.39$ rad, it is the largest systematic in our experiment. Of equal importance is the role this effect plays on other systematics, as we discuss below.

To model the phase shift due to parasitic lines, we follow the prescription outlined in \Refs \citenum{Carraz2012} and \citenum{Templier2021}. The PM generates a comb of laser lines with an electric field $E(t) = E_0 e^{i \omega_0 t} \sum_n i^n J_n(\beta) e^{i n\Delta\omega t}$, where $J_n(\beta)$ is a Bessel function with modulation depth $\beta$, and the modulation frequency $\Delta\omega = 2\pi \times 6.834$ GHz is the hyperfine splitting in $^{87}$Rb. Hence, each nearest-neighbour pair of lines is simultaneously resonant with a counter-propagating Raman transition. The Raman coupling parameter associated with each pairs of lines (one traveling upward, and its conjugate traveling downward) is proportional to the product of their electric field amplitudes, and inversely proportional to the detuning:
\be
  \Lambda_n \sim \frac{E_{n}(t) E_{n+1}^*(t - 2z/c)}{\Delta_{\rm Rb} + n\Delta\omega},
\ee
where $E_n(t) = E_0 e^{i \omega_0 t} i^n J_n(\beta) e^{i n\Delta\omega t}$ is an electric field amplitude, $z$ is the distance between the atoms and the mirror at time $t$, and $2z/c$ is the round-trip time required for the light to reflect off the mirror. It follows that
\begin{subequations}
\begin{align}
  \Lambda_n & \sim e^{i(k_{\rm Rb} z - \Delta\omega t)} \Omega_n e^{i n\Delta k z}, \\
  \Omega_n & \propto \frac{J_n(\beta) J_{n+1}(\beta)}{\Delta_{\rm Rb} + n\Delta\omega},
\end{align}
\end{subequations}
where $\Omega_n$ is a Rabi frequency and the effective wavevector is $k_{\rm Rb} = (2\omega_0 + \Delta\omega)/c$. The first term in $\Lambda_n$ describes the energy ($\hslash \Delta\omega$) and momentum ($\hslash k_{\rm Rb}$) transferred to the $^{87}$Rb atoms by the principal Raman lines during each pulse. These lines are associated with the principal Rabi frequency $\Omega_{0} \equiv \pi/2\tau$, where $\tau$ is the $\pi/2$-pulse duration. For other pairs of lines, the energy is identical, but due to additional spatial harmonics present in the field, the momentum transfer is modified to $\hslash(k_{\rm Rb} + n\Delta k)$, where $\Delta k = 2\Delta\omega/c$. This slightly different momentum kick for each pair of laser lines is the origin of the parasitic phase shift in the atom interferometer.

In the presence of additional laser lines, the effective Rabi frequency $\Omega_{\rm Rb}^{\rm eff}$ is given by the sum over all Rabi frequencies---each coupled with a phase term describing the modified momentum transfer
\be
  \label{OmegaRb(z)}
  \Omega_{\rm Rb}^{\rm eff}(z) = \sum_{n \in \mathbb{Z}} \Omega_{n} e^{i n \Delta k z}.
\ee
This results in a spatially-varying Rabi frequency due to the interference between different spatial harmonics. During each Raman pulse, the atoms are imprinted with the following phase:
\be
  \varphi(z) = {\rm arg}\left( \Omega_{\rm Rb}^{\rm eff}(z) \right).
\ee
In a three-pulse Mach-Zehnder interferometer, the total phase shift due to parasitic laser lines is
\be
  \phi_{\rm Rb}^{\ud, \rm PL} = \varphi(z_{\rm A}^{\ud}) - \varphi(z_{\rm B}^{\ud}) - \varphi(z_{\rm C}^{\ud}) + \varphi(z_{\rm D}^{\ud}),
\ee
where $z_{\rm A}, \ldots, z_{\rm D}$ denote the vertices of the interferometer at times $t = {\rm TOF}$, ${\rm TOF} + T$, and ${\rm TOF} + 2T$:
\begin{subequations}
\label{AIVertices}
\begin{align}
  z_{\rm A}^{\ud} & = -z_{\rm M} + v_{\rm Rb}^{\ud} {\rm TOF} + \frac{1}{2} g ({\rm TOF})^2, \\
  z_{\rm B}^{\ud} & = z_{\rm A}^{\ud} + (v_{\rm Rb}^{\ud} + g {\rm TOF}) T + \frac{1}{2} g T^2, \\
  z_{\rm C}^{\ud} & = z_{\rm A}^{\ud} + (v_{\rm Rb}^{\ud} + g {\rm TOF} \pm v_{\rm Rb}^{\rm rec}) T + \frac{1}{2} g T^2, \\
  z_{\rm D}^{\ud} & = z_{\rm A}^{\ud} + \left(v_{\rm Rb}^{\ud} + g {\rm TOF} \pm \frac{1}{2} v_{\rm Rb}^{\rm rec} \right) (2T) + \frac{1}{2} g (2T)^2.
\end{align}
\end{subequations}
Here, $v_{\rm Rb}^{\ud}$ is the initial velocity of the atomic cloud, $v_{\rm Rb}^{\rm rec} = \hslash k_{\rm Rb}/M_{\rm Rb}$ is the rubidium recoil velocity, and $z_{\rm M}$ is the position of the reference mirror.

\begin{figure}[!tb]
  \centering
  \includegraphics[width=0.48\textwidth]{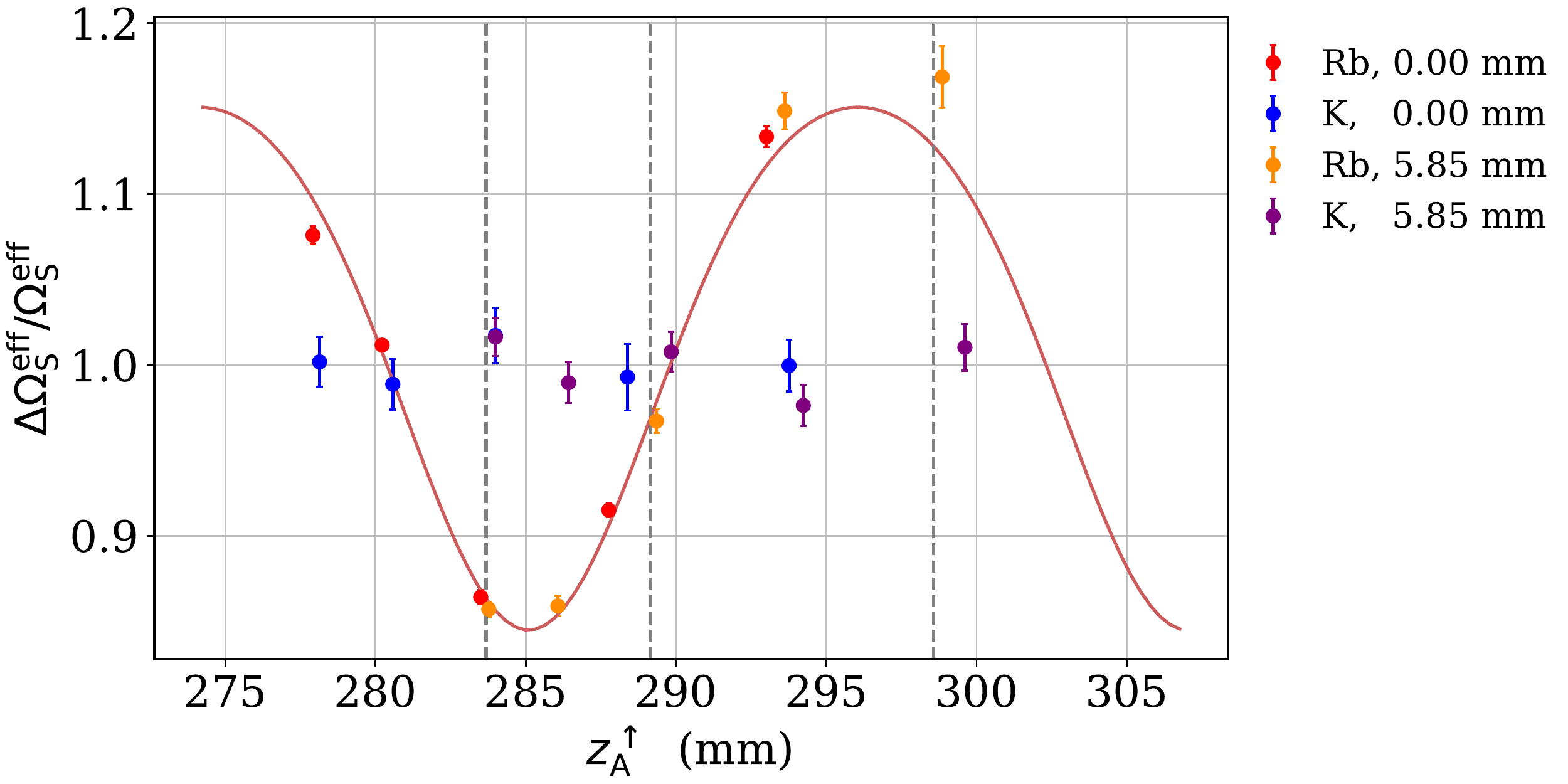}
  \caption{Relative change in the effective Rabi frequency $\Delta \Omega_{\rm S}^{\rm eff}/\Omega_{\rm S}^{\rm eff}$ measured at different positions relative to the reference mirror, where $\Omega_{\rm Rb}^{\rm eff} = 2\pi \times 98.4(9)$ kHz and $\Omega_{\rm K}^{\rm eff} = 2\pi \times 85.3(1.3)$ kHz. Two sets of data are shown for $^{87}$Rb and $^{39}$K: one at the nominal mirror position (red and blue points), and one with the mirror shifted upward by 5.85 mm (orange and purple points). The solid red curve is a fit to the two $^{87}$Rb data sets based on \Eq \eqref{OmegaRb(z)}. The vertical dashed lines indicate the position of the atoms during each interferometer pulse (TOF $= 16.2$ ms, $T = 20$ ms).}
  \label{fig:OmegavszM}
\end{figure}

The phase shift depends on the relative intensity of parasitic lines and the relative position between the atoms and the mirror. These parameters can be determined from the spatial variation of the Rabi frequency. Figure \ref{fig:OmegavszM} shows the relative change in Rabi frequency for both species as a function of the cloud position relative to the reference mirror $z_{\rm A}^{\uparrow}$. The Rabi frequency for $^{87}$Rb exhibits a strong spatial modulation due to parasitic lines, while $\Omega_{\rm K}^{\rm eff}$ is approximately constant. Here, the cloud position was controlled by varying the free-fall time before the Raman pulse. To extend the range of our measurements, we shifted the mirror position upward by 5.85 mm between data sets. Rabi frequencies were measured by varying the pulse duration and extracting $\Omega_{\rm S}^{\rm eff}$ from fits to the resulting Rabi oscillations. A fit to these data yields a nominal mirror position of $z_{\rm M} = -276.26(12)$ mm (the negative sign indicates the mirror is above the atoms), and a modulation depth of $\beta = 1.07(16)$. This is consistent with independent measurements of line intensities using a Fabry-Perot interferometer.

\begin{figure}[!tb]
  \centering
  \includegraphics[width=0.40\textwidth]{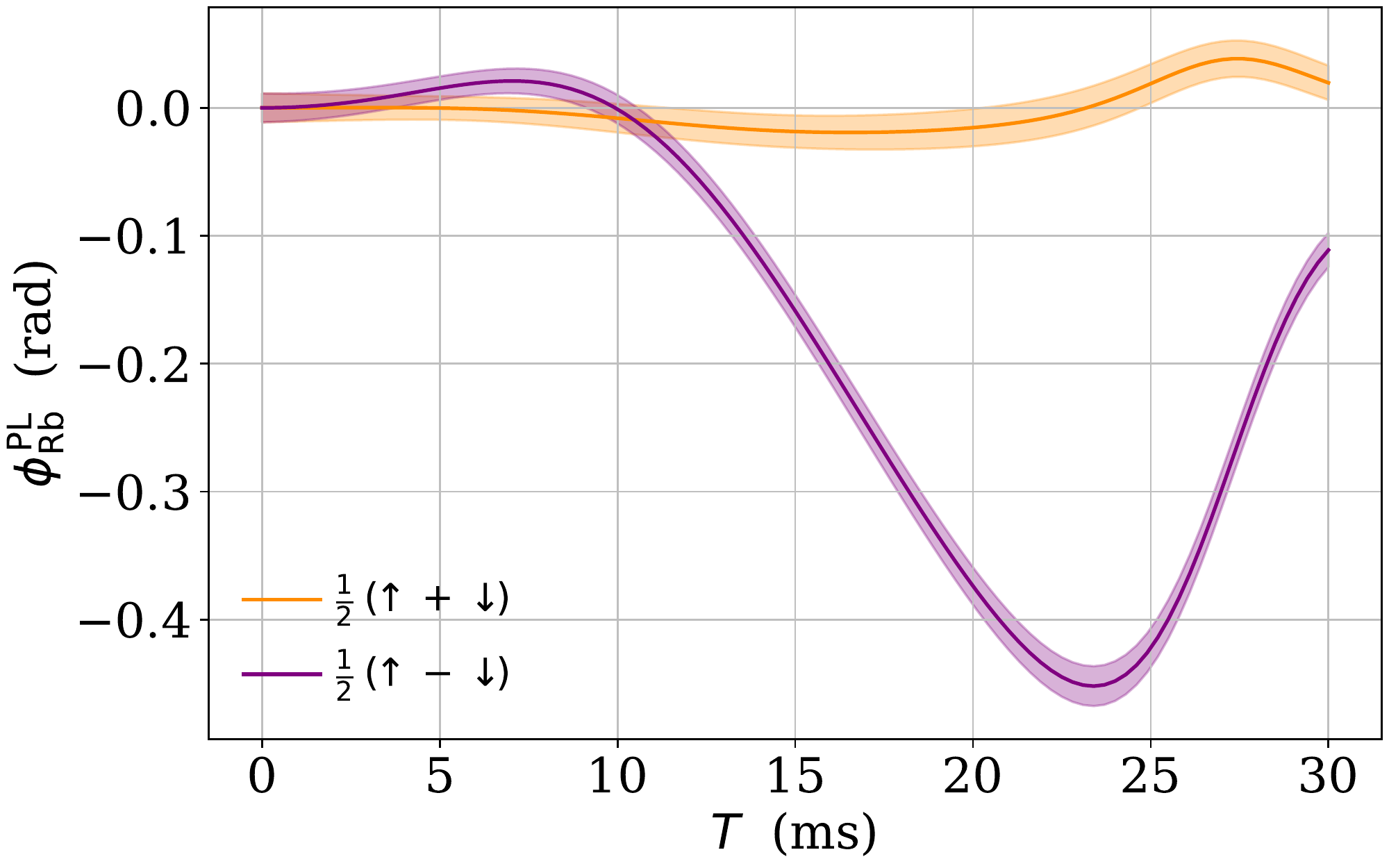}
  \caption{Phase shift due to parasitic laser lines as a function of the interrogation time $T$. The orange and purple curves represent the $k$-independent and $k$-dependent linear combinations of this systematic effect. The transparent regions indicate the $1\sigma$ confidence intervals given the uncertainty in our experimental parameters.}
  \label{fig:phiPLvsT}
\end{figure}

Using these parameters, we compute the phase shift as a function of $T$ in \Fig \ref{fig:phiPLvsT}. The phase has a dominant $k$-dependent contribution, due to the commonalities between the atomic trajectories, and hence it is not rejected with the $k$-reversal technique. The smaller $k$-independent component arises due to differences in the initial velocity and the sign reversal of the recoil term in \Eqs \eqref{AIVertices}. For the interrogation time used in our measurement ($T = 20$ ms) we estimate $\phi_{\rm Rb}^{\uparrow, \rm PL} = -0.394(27)$ rad and $\phi_{\rm Rb}^{\downarrow, \rm PL} = 0.378(27)$ rad.

\subsection{Asymmetry of the interferometers}

The effective Rabi frequency decreases in time due to the expansion of the cloud in the Raman beams. Although this effect is present in both species as a result of their finite temperature, it is completely dominated by the spatial modulation of the Rabi frequency caused by parasitic lines in $^{87}$Rb (see \Fig \ref{fig:OmegavszM}). The Rabi frequencies play a crucial role in the atom interferometer sensitivity function. Specifically, if they are not equal during the beamsplitter pulses, the Mach-Zehnder interferometer becomes asymmetric---causing velocity-dependent terms to emerge \cite{Bonnin2015}:
\be
  \label{phiAsym}
  \phi_{\rm S}^{\ud, \rm asym} = \mp k_{\rm S} \Delta v_{\rm S}^{\ud} \left[ \frac{1}{\Omega_{\rm S,3}^{\rm eff}} \tan\left(\frac{\Omega_{\rm S,3}^{\rm eff} \tau}{2}\right) - \frac{1}{\Omega_{\rm S,1}^{\rm eff}} \tan\left(\frac{\Omega_{\rm S,1}^{\rm eff} \tau}{2}\right) \right].
\ee
The term in brackets represents the asymmetry of the interferometer, and is non-zero only when there is an imbalance between the Rabi frequencies during the 1st and 3rd Raman pulses ($\Omega_{\rm S,1}^{\rm eff}$ and $\Omega_{\rm S,3}^{\rm eff}$). The velocity shift $\Delta_{\rm S}^{\ud}$ is the difference between velocity class addressed by the laser frequency and the true velocity of the atoms. This can arise due to the two-photon light shift, as well as the velocity-selectivity of the detection system. In our experiment, the latter is the dominant effect.

For $^{39}$K, we measure a small decrease in the relative Rabi frequency with the time-of-flight as a result of cloud expansion. We fit these data to a linear function $\Omega_{\rm K}^{\rm eff}(t) = \frac{\pi}{2\tau} (A t + B)$, and find a slope of $A = -3.6(5.9) \times 10^{-4}$ ms$^{-1}$ and an offset of $B = 1.013(22)$.

From these data, we estimate phase shifts of $\phi_{\rm Rb}^{\uparrow, \rm asym} = -63(72)$ mrad and $\phi_{\rm Rb}^{\uparrow, \rm asym} = 4(11)$ mrad. In terms of uncertainty, these shifts are limited by our knowledge of the initial velocity, the atom-mirror distance and the mean Rabi frequency for each species.

\subsection{Two-photon light shift}
\label{sec:TPLS}

The two-photon light shift (TPLS) is a velocity-dependent frequency shift of the counter-propagating Raman transition due to the presence of neighbouring off-resonant transitions. To compute the phase shift due to the TPLS, we follow the approach of \Ref \citenum{Gauguet2008}. Ignoring effects due to magnetically-sensitive co-propagating transitions, the TPLS is composed of two contributions
\be
  \label{TPLS-Simple}
  \omega_{\rm S}^{\ud, \rm TPLS}(t) = \omega_{\rm S}^{\ud, \rm  counter}(t) + \omega_{\rm S}^{\ud, \rm co}(t).
\ee
The first term is due to counter-propagating transitions:
\be
  \label{eq:TPLSContra}
  \omega_{\rm S}^{\ud,\rm counter}(t) = \frac{1}{4} \left( \frac{(\Omega_{\rm S}^{\rm eff})^2}{2 \omega_{\rm S}^{\ud, \rm D}} + \frac{(\Omega_{\rm S}^{\rm eff})^2}{2\omega_{\rm S}^{\ud, \rm D} + 4\omega_{\rm S}^{\rm rec}} \right).
\ee
with effective Rabi frequency $\Omega_{\rm S}^{\rm eff}$ and recoil frequency $\omega_{\rm S}^{\rm rec} = \hslash k_{\rm S}^{2}/2 M_{\rm S}$. This shift explicitly depends on the time during the interferometer and the momentum transfer direction due to the Doppler shift $\omega_{\rm S}^{\ud, \rm D} = \mp k_{\rm S} \big(v_{\rm S}^{\ud} + g (t-\mbox{TOF})\big)$. The second term in \Eq \eqref{TPLS-Simple} is due to residual co-propagating transitions between $|F=1, m_F=0\rangle$ and $|F=2, m_F=0\rangle$
\be
  \label{eq:TPLSCo}
  \omega_{\rm S}^{\ud,\rm co}(t) = \frac{(\Omega_{\rm S}^{\rm co})^2}{4 (\omega_{\rm S}^{\ud, \rm D} + \omega_{\rm S}^{\rm rec})}.
\ee
This term arises due to imperfect crossed-linear polarization of the Raman beams. We measure the Rabi frequencies $\Omega_{\rm S}^{\rm eff}$ and $\Omega_{\rm S}^{\rm co}$ by addressing the corresponding resonance and measuring the Rabi oscillations that result from varying the Raman pulse duration. For the counter-propagating Rabi frequencies, we find mean values of $\Omega_{\rm Rb}^{\rm eff} = 2\pi \times 98.4(9)$ kHz and $\Omega_{\rm K}^{\rm eff} = 2\pi \times 85.3(1.3)$ kHz over the duration of the interferometers. These Rabi frequencies vary due to the effects of parasitic lines (in $^{87}$Rb) and cloud expansion. We model these effects an include them in our estimates of the TPLS. Similarly, we find residual co-propagating Rabi frequencies of $\Omega_{\rm Rb}^{\rm co} \simeq 0.20~\Omega_{\rm Rb}^{\rm eff}$ and $\Omega_{\rm K}^{\rm co} \simeq 0.28~\Omega_{\rm K}^{\rm eff}$.

The phase shift on the atom interferometer can be derived from the sensitivity function:
\be 
  \phi_{\rm S}^{\ud,\rm TPLS} = \int g_{\rm S}(t) \omega_{\rm S}^{\ud,\rm TPLS}(t) \dd t.
\ee
Assuming the TPLS is constant during the Raman pulses, the phase shift reduces to
\be 
  \label{phiTPLS}
  \phi_{\rm S}^{\ud, \rm TPLS} = \frac{\omega_{\rm S,3}^{\ud,\rm TPLS}}{\Omega_{\rm S,3}^{\rm eff}} \tan\left( \frac{\Omega_{\rm S,3}^{\rm eff} \tau}{2} \right) - \frac{\omega_{\rm S,1}^{\ud,\rm TPLS}}{\Omega_{\rm S,1}^{\rm eff}} \tan\left( \frac{\Omega_{\rm S,1}^{\rm eff} \tau}{2} \right),
\ee
where $\Omega_{\rm S,i}^{\rm eff}$ is the effective Rabi frequency during the $i$th pulse. The different contributions to the TPLS phase shift are summarized in Table \ref{tab:TPLS}.

\begin{table}[!tb]
  \caption{Top rows: two-photon light shifts for each species during the 1st and 3rd Raman pulses at $t = 16.2$ ms and 56.2 ms, respectively. Bottom row: phase shifts due to the TPLS. Other parameters: $v_{\rm Rb}^{\uparrow} = 1.74(69)$ cm/s, $v_{\rm K}^{\uparrow} = 3.07(60)$ cm/s, $v_{\rm Rb}^{\downarrow} = 1.64(65)$ cm/s, $v_{\rm K}^{\downarrow} = 2.63(51)$ cm/s, $z_{\rm M} = -276.26(12)$ mm, and $\beta = 1.07(16)$.}
  \label{tab:TPLS}
  \begin{ruledtabular}
  \begin{tabular}{cccccc}
    Pulse $i$ & $\Omega_{{\rm Rb}, i}^{\uparrow, \rm TPLS}$ & $\Omega_{{\rm Rb}, i}^{\downarrow, \rm TPLS}$ & $\Omega_{{\rm K}, i}^{\uparrow, \rm TPLS}$ & $\Omega_{{\rm K}, i}^{\downarrow, \rm TPLS}$ & Units \\
    \hline
     1 & $-4.4(1.8)$ & $4.1(1.7)$ & $-4.20(26)$  & $3.72(21)$  & kHz \\
     3 & $-2.16(69)$ & $2.20(69)$ & $-1.230(65)$ & $1.238(65)$ & kHz \\
    \hline
       & $\phi_{\rm Rb}^{\uparrow, \rm TPLS}$ & $\phi_{\rm Rb}^{\downarrow, \rm TPLS}$ & $\phi_{\rm K}^{\uparrow, \rm TPLS}$ & $\phi_{\rm K}^{\downarrow, \rm TPLS}$ &  \\
    \hline
       & $17(18)$ & $-14(18)$ & $27.7(2.5)$ &  $-23.2(2.1)$ & mrad \\
  \end{tabular}
  \end{ruledtabular}
\end{table}
\subsection{One-photon light shift}
\label{sec:OPLS}

The one-photon light shift (OPLS), also known as the differential AC Stark shift, is a shift between the ground states due to the far off-resonant Raman beams. In this case, the shift on the ground state $\ket{F}$ scales as the ratio of optical intensity $I_j$ over the detuning $\Delta_{F',F,j} = \omega_j - (\omega_{F'} - \omega_F)$ from each excited state $\ket{F'}$ \cite{Hu2017}. The total OPLS is then the sum of the shifts produced by each optical field $j$:
\be
  \omega^{\rm OPLS} = \sum_{F,F',j} \frac{|\langle F || \bm{d} || F' \rangle|^2}{6 \hslash^2 c \epsilon_0} \frac{I_j}{\Delta_{F',F,j}}.
\ee
where $|\langle F || \bm{d} || F' \rangle|$ is a reduced dipole matrix element, $c$ is the speed of light and $\epsilon_0$ is the vacuum permittivity. Here, we have included only effects due to the scalar polarizability \cite{Hu2017}. For rubidium, the OPLS is strongly suppressed by operating at the correct intensity ratio between the two beams. Accounting for all lines produced by the phase modulator, this ratio is $I_2/I_1 \simeq 1.71$ for a Raman detuning of $\Delta_{\rm Rb} \equiv \Delta_{2',2,2} = -1.1$ GHz. For potassium, because the hyperfine splitting is much smaller than the Raman detuning, it is not possible to suppress the OPLS by adjusting the intensity ratio. Instead, we operate the interferometer with as large a detuning as possible ($\Delta_{\rm K} \equiv \Delta_{2',2,2} = -1.34$ GHz) to minimize the shift. In this configuration, however, the OPLS is much stronger in $^{39}$K than $^{87}$Rb. Fortunately, the geometry of Mach-Zehnder interferometer rejects constant frequency shifts since atoms spend an equal amount of time in each ground state. As a result, we are primarily sensitive to \emph{variations} in the OPLS during the interferometers. Due to the due to the Gaussian spatial profile of the Raman beams, and the expansion of the clouds in the beams, the ``mean'' OPLS varies differently for each species.

We measure the OPLS by performing Raman spectroscopy with counter-propagating beams. A complete set of spectra for different laser intensities and free-fall times allows us to extract the frequency shifts during each interferometer. We observe a linear variation of this shift over 60 ms of free-fall time. Table \ref{tab:OPLS} summarizes the one-photon light shifts for each species during the first and third Raman pulses. We estimate the phase shift due to the OPLS using the sensitivity function
\be 
  \label{phiOPLS}
  \phi_{\rm S}^{\rm OPLS} = \frac{\omega_{\rm S,3}^{\rm OPLS}}{\Omega_{\rm S,3}^{\rm eff}} \tan\left( \frac{\Omega_{\rm S,3}^{\rm eff} \tau}{2} \right) - \frac{\omega_{\rm S,1}^{\rm OPLS}}{\Omega_{\rm S,1}^{\rm eff}} \tan\left( \frac{\Omega_{\rm S,1}^{\rm eff} \tau}{2} \right),
\ee
where $\omega_{\rm S, i}^{\rm OPLS}$ is the frequency shift of the counter-propagating resonance during the $i$th Raman pulse. From these data, we estimate $\phi_{\rm Rb}^{\rm OPLS} = 6.2(6.6)$ mrad and $\phi_{\rm K}^{\rm OPLS} = -7.2(4.7)$ mrad.

\begin{table}[!tb]
  \caption{One-photon light shift for each species during the first and third Raman pulses at times $t =$ TOF and $t = \mbox{TOF} + 2T$.}
  \label{tab:OPLS}
  \begin{ruledtabular}
  \begin{tabular}{cccc}
    Pulse $i$ & Time (ms) & $\omega_{{\rm Rb},i}^{\rm OPLS}$ (kHz) & $\omega_{{\rm K},i}^{\rm OPLS}$ (kHz) \\
  \hline
    1 & 16.2 & -4.4(2.9) & -105.6(2.1) \\
    3 & 56.2 & -0.3(2.9) & -108.5(2.1) \\
  \end{tabular}
  \end{ruledtabular}
\end{table}

As this phase shift is identical for both momentum transfer directions, we reject it using the $k$-reversal technique. However, this rejection is imperfect because of fluctuations in the laser intensity, atom cloud size and temperature. These effects average to zero over many measurements, but result in a non-zero uncertainty of $\sim 3$ mrad.

\subsection{Beam misalignments}
\label{sec:Misalignments}

The measurement of each atom's acceleration is subject to errors in the Raman wavevector's magnitude and direction through the scalar product $\bm{k}_{\rm S} \cdot \bm{a}_{\rm S} T_{\rm eff}^2$. These errors contribute a systematic phase shift
\be
  \label{phibeam}
  \phi_{\rm S}^{\ud, \rm beam} = \pm \left( \frac{2\Delta \omega_{\rm S}^{\rm las}}{\omega_{\rm S}^{\rm las}} - \frac{1}{2} \theta_{\rm S}^2 \right) k_{\rm S} g T_{\rm eff}^2,
\ee
where $\Delta \omega_{\rm S}^{\rm las}$ is the frequency error between the measured Raman laser frequency and its true value, and $\theta_{\rm S}$ is a small misalignment angle of the Raman beam relative to vertical. We have calibrated the absolute frequency of both Raman lasers at the level of $\sim 500$ kHz using saturated absorption spectroscopy and an optical frequency comb. This error represents a phase shift of $< 200$ $\mu$rad for both species, hence the 1st term of \Eq \eqref{phibeam} is negligible. The 2nd term, however, can be large since it scales as $\theta_{\rm S}^2$ due to the $\cos\theta_{\rm S}$ that appears in the scalar product. We emphasize that only the relative angle $\Delta \theta = \theta_{\rm Rb} - \theta_{\rm K}$ between the $^{87}$Rb and $^{39}$K Raman beams is important for the measurement of $\eta$. This relative angle can only be produced by differences in alignment between the incident beams, as the retro-reflection mirror is common to both species. Corrections to $\eta$ caused by the non-verticality of the mean acceleration $\frac{1}{2}(\bm{a}_{\rm K} + \bm{a}_{\rm Rb})$ (\ie due to the reference mirror) can be safely ignored since they scale as $\Delta \theta^4$.

To measure the beam misalignment, we imaged the Raman beams on two planes separated by a distance of $\sim 5$ m. The separation between the beam centers yielded a relative angle of $\Delta \theta = 2.80(30)$ mrad---limited by our knowledge of the beam centers and the distance between the two planes. Taking the $^{39}$K Raman beam as our reference ($\theta_{\rm K} \equiv 0$, $\theta_{\rm Rb} = \Delta\theta$), the phase shift on the $^{87}$Rb interferometer is $\phi_{\rm Rb}^{\uparrow, \rm beam} = -0.248(53)$ rad. This misalignment originates from a commercial collimator, where the two Raman beams arrive by optical fiber and are combined in free space. The relatively large misalignment angle was discovered post measurement, and can be significantly reduced in future measurements with simple adjustments.


\subsection{Wavefront aberration and curvature}
\label{sec:Wavefront}

Imperfections in the optics along the Raman beams create aberrations that distort their wavefronts. Differences between the incident and reflected Raman wavefronts creates a phase inhomogeneity over the atom cloud as it expands. If the average value of the phases imprinted on the cloud is not zero, these imperfections lead to a systematic shift. This effect is related to the ballistic expansion of the atomic source, and cancels at zero temperature. Nevertheless, it has been shown to be very challenging to extrapolate the measurement to zero temperature \cite{Louchet-Chauvet2011a, Trimeche2017, Karcher2018}. Moreover, we emphasize that there is little to no rejection of wavefront distortions between the two atom interferometers because of differences in mass, temperature, and initial cloud size.

We first focus on the phase shift due to wavefront curvature (WC), which is determined by the collimation of the Raman beams and the temperature of the sample \cite{Louchet-Chauvet2011a, Trimeche2017}
\be
  \label{WavefrontCurvature}    
  \phi_{\rm S}^{\ud, \rm WC} = \pm k_{\rm S} \frac{(\sigma_{\rm S}^{\rm vel} T)^2}{R_{\rm S}}.
\ee
Here, $\sigma_{\rm S}^{\rm vel} = \sqrt{k_{\rm B} \mathbb{T}_{\rm S}/M_{\rm S}}$ is the rms width of the velocity distribution, and $R_{\rm S}$ is the radius of curvature of the Raman beam. The temperature of each species is determined using Raman spectroscopy to be $\mathbb{T}_{\rm Rb} = 5.0(5)$ $\mu$K and $\mathbb{T}_{\rm K} = 4.7(5)$ $\mu$K, respectively. The radius of curvature is estimated from the divergence of the beams, which is determined using the two aforementioned Raman beam images. The beam profiles are fit in 2D to obtain their beam waists at two positions. The two Raman beams are not identically collimated: the Rayleigh length for $^{87}$Rb is $z_{\rm Rb}^{\rm R} = 113.08(10)$ m, and $z_{\rm K}^{\rm R} = 23.39(10)$ m for $^{39}$K. This produces radii of curvature $R_{\rm Rb} \simeq (z_{\rm Rb}^{\rm R})^2/z = 16350(32)$ m and $R_{\rm K} = 700.3(6.7)$ m at the location of the atoms (a distance of $z = 0.78(10)$ m from the collimator). The resulting phase shifts due to wavefront curvature are $\phi_{\rm Rb}^{\uparrow, \rm WC} = 0.189(38)$ mrad and $\phi_{\rm K}^{\uparrow, \rm WC} = 9.39(20)$ mrad.

Wavefront aberrations (WA) due to the various optics (viewports, quarter waveplate, reference mirror) are difficult to measure at the location of the atoms. To estimate the effect, we measured the wavefront profile of the Raman beams with a Shack-Hartmann analyser (Thorlabs WFS20-5C), at different positions: before the vacuum system, after a first path through the vacuum chamber, and after the retro-reflection mirror and a second pass through chamber. We did not observe a significant distortion of the wavefront because our measurement was limited by the resolution of the Shack-Hartmann analyser. Nevertheless, the output of the analyser gives a lower-bound on the residual radius of curvature: $R_{\rm min} = 46.6$ m. This corresponds to a maximum peak-to-valley distortion of 31 nm over a measurement diameter of 3.4 mm. We do not assign a shift due to wavefront aberrations. Instead, we use the minimum curvature to estimate uncertainties of $\delta \phi_{\rm Rb}^{\ud, \rm WA} = 0.066$ rad and $\delta \phi_{\rm K}^{\ud, \rm WA} = 0.141$ rad. Due to the relatively large velocity dispersion of the potassium sample, this effect is the most significant contribution to our error budget.




\subsection{Coriolis force}
\label{sec:Coriolis}

For an atom with an initial velocity in the horizontal plane, the interferometer arms enclose a spatial area, which makes it sensitive to rotations via the Sagnac effect \cite{Louchet-Chauvet2011a}. For an initial atomic velocity $v_{\rm S}^x$ along the $x$-axis (pointing westward), an additional phase shift due to the Coriolis effect of Earth's rotation leads to a bias for each species
\be
  \phi_{\rm S}^{\ud, \rm Cor}
  = \pm \bm{k}_{\rm S} \cdot \bm{a}_{\rm Cor} T^2
  = \mp 2 k_{\rm S} v_{\rm S}^x \Omega_{\bigoplus} \cos(\vartheta) T^2,
  \label{eq:Coriolis}
\ee
where $\bm{a}_{\rm Cor} = 2\bm{v} \times \bm{\Omega}$ is the Coriolis acceleration, $\bm{\Omega} = (0, -\Omega_{\bigoplus} \cos\vartheta, -\Omega_{\bigoplus} \sin\vartheta)$ is the Earth's rotation vector in the lab frame, and $\Omega_{\bigoplus} \cos\vartheta = 5.17 \times 10^{-5}$ rad/s at our latitude of $\vartheta = 44.8$ deg. To estimate this effect, we measured the initial transverse velocity of the atoms in the laboratory frame using time-of-flight imaging on with a calibrated CCD camera. We measured $v_{\rm Rb}^x = -1.0(1)$ mm/s and $v_{\rm K}^x = -4.0(4)$ mm/s, which are due to a slight misalignment and power imbalance between the MOT beams. We estimate $\phi_{\rm Rb}^{\uparrow,\rm Cor} = 0.67(7)$ mrad and $\phi_{\rm K}^{\uparrow,\rm Cor} = 2.7(3)$ mrad for an interrogation time of $T = 20$ ms.

\subsection{Magnetic gradient force}
\label{sec:MagneticGradient}

For atoms in the magnetically-insensitive $m_F = 0$ state, the magnetic field produces a potential of the form $U_{\rm S} = \mp \frac{1}{2} h K_{\rm S} B^2(z)$ due to the 2nd-order Zeeman effect, where the sign is negative (positive) for atoms in the $\ket{F=1}$ ($\ket{F = 2}$) ground state. Hence, a gradient in the magnetic field produces a force on the atoms given by
\be
  \mathbb{F}_{\rm S}^{\ket{1},\ket{2}} = \pm h K_{\rm S} B(z) \nabla B_z \simeq \pm h K_{\rm S} (B_0 + \nabla B_z \cdot z) \nabla B_z,
\ee
where $h$ is Planck's constant, and $K_{\rm S}$ is given by \Eq \eqref{KS}. Since the force is opposite in sign for the two internal states, and the atom spends half the time in each state during the interferometer, the contribution from the constant term $B_0 \nabla B_z$ cancels in the atom interferometer phase shift. However, the term proportional $(\nabla B_z)^2 z$ is non-zero due to the asymmetric sampling of magnetic field during the atom's free-fall trajectory. Up to order $T^4$ and $\Lambda_{\rm S}^2 = h K_{\rm S}/M_{\rm S}$, the phase shift arising from this state-dependent force can be shown to be \cite{Barrett2016}
\be
  \phi_{\rm S}^{\ud,\nabla B} = \mp \frac{2}{3} k_{\rm S} \big(\Lambda_{\rm S} \nabla B_z \big)^2 \left[\left(v_{\rm S,1}^{\ud} \pm \frac{1}{2} v_{\rm S}^{\rm rec} \right) T + g T^2 \right] T^2,
\ee
where $v_{\rm S,1}^{\ud}$ is the atomic velocity at the first Raman pulse. For our experimental parameters, we estimate $\phi_{\rm Rb}^{\uparrow, \rm MGF} = -0.118(7)$ mrad and $\phi_{\rm K}^{\uparrow, \rm MGF} = -4.18(24)$ mrad.

We emphasize that this phase shift is $(\Lambda_{\rm K}/\Lambda_{\rm Rb})^2 \sim 33$ times larger for $^{39}$K than $^{87}$Rb due to its lighter mass and smaller hyperfine splitting. It will therefore be an important effect to consider in future measurements with large interrogation times.

\subsection{Gravity gradient}
\label{sec:GravityGradient}

A linear gravity gradient will modify the atom's free-fall trajectory relative to a parabola---causing a phase shift of the interferometer. This phase shift is given by \cite{Wolf1999, Storey1994}
\be
  \label{GravityGradient}
  \phi_{\rm S}^{\ud, \rm GG} = \pm k_{\rm S} T_{zz} \left[z_{\rm S,1} + \left(v_{\rm S,1}^{\ud} \pm \frac{1}{2} v_{\rm S}^{\rm rec} \right) T + \frac{7}{12} g T^2 \right] T^2,
\ee
where $z_{\rm S,1}$ is the cloud position at the first Raman pulse, $v_{\rm S,1}^{\ud}$ is the velocity at the first Raman pulse, and $T_{zz} \simeq +3.1 \times 10^{-6}$ s$^{-2}$ is the vertical gravity gradient of the Earth (positive indicates increasing downward). This effect is proportional to $k_{\rm S}$ and cannot be rejected using the $k$-reversal method. However, the phase shift is negligible in our case because of our modest interrogation time: $\phi_{\rm Rb}^{\uparrow, \rm GG} = 149(10)$ $\mu$rad and $\phi_{\rm K}^{\uparrow, \rm GG} = 164(10)$ $\mu$rad. The uncertainty in these estimates arises from the error in our knowledge of the initial cloud position ($\delta z_{\rm S,1} \sim 0.3$ mm) and the initial velocity ($\delta v_{\rm S,1}^{\ud} \sim 6$ mm/s). The latter is dominated by our calibration of the detection system, which is moderately velocity selective (see \Sec \ref{sec:AtomicVelocity}).

The gravity gradient effect can dominate for larger values of $T$, where differences in initial cloud position and velocity play a strong role. A method to cancel the gravity gradient phase shift was recently proposed in \Ref \citenum{Roura2014}, which involves modifying the norm of the $k$-vector during the $\pi$-pulse. This technique has also been demonstrated experimentally \cite{DAmico2017, Overstreet2018}.

\subsection{Total uncertainty on the E\"{o}tv\"{o}s parameter}

Table \ref{tab:IndividualSystematics} presents a summary of the systematic shifts and their uncertainties for each species and momentum transfer direction. The shifts are listed in order of significance in terms of their contribution to the total uncertainty budget. The $^{39}$K interferometer has no shift due to beam misalignment because we use the $^{39}$K Raman beam as a reference to measure the misalignment with the $^{87}$Rb Raman beam. Similarly, there is no shift due to parasitic lines because there is no phase modulator present in the $^{39}$K Raman laser system.

\begin{table*}[!tb]
  \caption{Systematic shifts on measurements of the gravitational acceleration for each atomic species and momentum transfer direction. Values are expressed as accelerations $a_{\rm S}^{\ud, \rm sys} = \phi_{\rm S}^{\ud, \rm sys}/k_S T_{\rm eff}^2$, and uncertainties are shown in parentheses. Experimental parameters: $\mbox{TOF} = 16.2$ ms, $T = 20$ ms, $\tau = 2.5$ $\mu$s, $z_{\rm S}^0 = 0.0(0.5)$ mm, $v_{\rm Rb}^{\uparrow} = 1.74(69)$ cm/s, $v_{\rm K}^{\uparrow} = 3.07(60)$ cm/s, $v_{\rm Rb}^{\downarrow} = 1.64(65)$ cm/s, $v_{\rm K}^{\downarrow} = 2.63(51)$ cm/s, $B_0 = 143.24(78)$ mG, $\nabla B_z = -1.172(33)$ G/m, $A_{\rm Eddy} = -52.8(1.0)$ mG, $\Gamma_{\rm Eddy} = 50.25(38)$ s$^{-1}$, $z_{\rm M} = -276.26(12)$ mm, and $\beta = 1.07(16)$.}
  \label{tab:IndividualSystematics}
  \setlength\tabcolsep{0.pt}
  \begin{ruledtabular}
  \begin{tabular}{cdddd}
    Systematic effect &
    \multicolumn{1}{r}{$a^{\uparrow, \rm sys}_{\rm Rb}$ ($\times 10^{-6}~g$)} & \multicolumn{1}{r}{$a^{\downarrow, \rm sys}_{\rm Rb}$ ($\times 10^{-6}~g$)} &
    \multicolumn{1}{r}{$a^{\uparrow, \rm sys}_{\rm K}$ ($\times 10^{-6}~g$)} & \multicolumn{1}{r}{$a^{\downarrow, \rm sys}_{\rm K}$ ($\times 10^{-6}~g$)} \\
    \hline
      Wavefront aberration   &  0.0(1.1)    &  0.0(1.1)    &  0.0(2.2)    &   0.0(2.2)    \\
      2nd-Order Zeeman       &  0.097(95)   &  0.181(93)   & -0.6(1.4)    &   2.4(1.4)  \\
      AI asymmetry           & -1.0(1.1)    &  1.0(1.1)    &  0.06(16)    &  -0.05(14)    \\
      Beam misalignment      & -3.92(84)    &  3.92(84)    &  0           &   0           \\
      Parasitic lines        & -6.24(43)    &  5.98(42)    &  0           &   0           \\
      Two-photon light shift &  0.27(29)    & -0.23(28)    &  0.431(39)   &  -0.361(33)   \\
      One-photon light shift &  0.10(10)    &  0.10(10)    & -0.103(68)   &  -0.103(68)   \\
      Wavefront curvature    &  0.00298(60) & -0.00298(60) &  0.146(31)   &  -0.146(31)   \\
      Coriolis force         &  0.0106(11)  & -0.0106(11)  &  0.0422(42)  &  -0.0422(42)  \\
      Magnetic force         & -0.00187(11) &  0.00180(11) & -0.0650(38)  &   0.0599(35)  \\
      Gravity gradient       &  0.00236(15) & -0.00228(15) &  0.00256(16) &  -0.00234(15) \\
    \hline
      Total systematics      & -10.7(1.8)   & 10.9(1.8)    & -0.1(2.6)    &  1.7(2.6)     \\
  \end{tabular}
  \end{ruledtabular}
\end{table*}

Table \ref{tab:EtaSystematics} summarizes the total uncertainty budget for the E\"{o}tv\"{o}s parameter. Here, each row is computed from Table \ref{tab:IndividualSystematics} using the following linear combination of $\pm k_{\rm S}$ to reject the $k$-independent effects:
\be
  \label{etasys}
  \eta^{\rm sys} = \frac{1}{g} \left[\frac{a_{\rm K}^{\uparrow, \rm sys} - a_{\rm K}^{\downarrow,\rm sys}}{2} - \frac{a_{\rm Rb}^{\uparrow, \rm sys} - a_{\rm Rb}^{\downarrow,\rm sys}}{2} \right].
\ee
The uncertainty due to each systematic $\delta\eta^{\rm sys}$ is computed as the quadratic sum of the $k$-dependent systematic uncertainties for each species 
\be
  \label{detasys}
  (\delta \eta^{\rm sys})^2 = \frac{1}{g^2} \left[ (\delta a_{\rm K}^{\uparrow - \downarrow, \rm sys})^2 + (\delta a_{\rm Rb}^{\uparrow - \downarrow, \rm sys})^2 \right].
\ee
This assumes there is inter-species correlation between systematic effects. However, we allow for a correlation between the momentum transfer directions within each species. As a conservative estimate, we consider that they are correlated at the 50\% level (\ie the correlation coefficient between $\pm k_{\rm S}$ systematics is $R_{\rm S}^{\rm sys} = 0.50$). The resulting uncertainty in the $k$-dependent systematic shift is
\begin{align}
  (\delta a_{\rm S}^{\uparrow-\downarrow, \rm sys})^2
  & = \frac{1}{4} \left[ (\delta a_{\rm S}^{\uparrow, \rm sys})^2 + (\delta a_{\rm S}^{\downarrow, \rm sys})^2 \right. \\
  & \;\;\;\;\;\;\;\;\;\;\; \left. - \; 2 R_{\rm S}^{\rm sys} \delta a_{\rm S}^{\uparrow, \rm sys} \delta a_{\rm S}^{\downarrow, \rm sys} \right]. \nonumber
\end{align}

We find good agreement between our raw measurement $\eta^{\rm raw} = 10.79(8) \times 10^{-6}$ and the estimated total systematic shift $9.9(1.6) \times 10^{-6}$. Our final measurement of the E\"{o}tv\"{o}s parameter is therefore $\eta = 0.9(1.6) \times 10^{-6}$, which is limited entirely by systematic effects.

\begin{table}[!tb]
  \caption{Systematic shifts on the E\"{o}tv\"{o}s parameter and their uncertainties in order of significance. The final measurement is the difference between the raw measurement and the sum of all systematic shifts.}
  \label{tab:EtaSystematics}
  \begin{ruledtabular}
  \begin{tabular}{cdd}
    Systematic effect & \multicolumn{1}{c}{$\eta$ ($\times 10^{-6}$)} & \multicolumn{1}{c}{$\delta\eta$ ($\times 10^{-6}$)} \\
    \hline
    Wavefront aberration   &  0.00  &  1.21  \\
    2nd-Order Zeeman       & -1.44  &  0.69  \\
    AI asymmetry           &  1.05  &  0.56  \\
    Beam misalignment      &  3.92  &  0.42  \\
    Parasitic lines        &  6.11  &  0.19  \\
    Two-photon light shift &  0.15  &  0.14  \\
    One-photon light shift & <0.01  &  0.06  \\
    Wavefront curvature    &  0.14  &  0.02  \\
    Coriolis force         &  0.03  & <0.01  \\
    Magnetic force         & -0.06  & <0.01  \\
    Gravity gradient       & <0.01  & <0.01  \\
    \hline
    Total systematics      &  9.90  &  1.58  \\
    Raw measurement        & 10.79  &  0.08  \\
    \hline
    Final measurement      &  0.89  &  1.58  \\
  \end{tabular}
  \end{ruledtabular}
\end{table}
\section{Conclusion}
\label{sec:Conclusion}


We present a test of the Universality of Free Fall with simultaneous $^{39}$K-$^{87}$Rb interferometers. The high level of correlation between these species enabled us to reach a state-of-the-art sensitivity on the E\"{o}tv\"{o}s parameter of $\delta \eta^{\rm stat} = 7.8 \times 10^{-8}$ after $\sim 7$ hours of integration with a modest interrogation time of $T = 20$ ms. We also evaluated the accuracy of our measurement through a detailed analysis of systematic effects. Our final measurement yielded $\eta = 0.9(1.6) \times 10^{-6}$, which is consistent with no violation of the UFF. Systematic shifts due to the 2nd-order Zeeman effect, wavefront distortions, parasitic lines in the rubidium laser, and misalignments between our Raman beams are the most significant contributions to our uncertainty budget. We also discovered a velocity shift introduced by our detection system, which contributed additional shifts due to a coupling with variation of the Rabi frequency over the duration the interferometers. Our work highlights specific challenges associated with utilizing atomic species with a large mass difference, and paves the way for more accurate tests with rubidium and potassium in the future.

In the near term, we will upgrade our experiment by using ultracold atoms in the microgravity environment provided by an Einstein elevator \cite{Condon2019}. We expect to gain several orders of magnitude on both sensitivity and accuracy with this setup, since the atoms can be interrogated for several hundred milliseconds in the same position relative to reference mirror. Using ultracold atoms will also drastically reduce the effects of wavefront distortions, which scale as the temperature of the sample \cite{Karcher2018}. In the long term, further improvements will be possible using large momentum transfer atom optics and squeezed states. 

The full potential of atom interferometers can only be realized in Space. In this context, our experiment serves as a low-cost engineering model tested in a relevant environment. Our work is an important step in the preparation for a quantum test of the Universality of Free Fall below $10^{-15}$ that will probe the frontier of General Relativity and Quantum mechanics \cite{Aguilera2014}. Finally, the improvement and validation of cold-atom technology used in our experiments is beneficial to other applications, such as Earth gravity surveys \cite{Menoret2018}, gravitational-wave detection \cite{Canuel2014}, and inertial navigation \cite{Cheiney2018}.

The data that support the findings of this study are available from the corresponding author upon reasonable request.



\section{Acknowledgments}

This work is supported by the French national agency CNES (Centre National d'Etudes Spatiales), and the European Space Agency (ESA). For financial support, B. Barrett, R. Arguel and L. Antoni-Micollier wish to thank CNES and IOGS; G. Condon and V. Jarlaud wish to thank IOGS; and C. Pelluet thanks CNES and ESA. Finally, P. Bouyer thanks Conseil R\'{e}gional d'Aquitaine for the Excellence Chair.

\appendix

\section{The differential FRAC signal}
\label{sec:FRACSignal}


In this Appendix, we derive the differential FRAC signal for a dual-species interferometer, and we discuss sources of error due to the classical sensor used to correct each interferometer phase. For convenience, we write the total phase shift each atom interferometer as four separate contributions: one from the inertial effects $\phi_{\rm S}^{\rm acc}$, one from the interrogation laser $\phi_{\rm S}^{\rm las}$, one from vibrations of the reference frame $\phi_{\rm S}^{\rm vib}$, and one from systematics $\phi_{\rm S}^{\rm sys}$. Using the sensitivity function formalism, the total phase is
\begin{subequations}
\label{phitot}
\begin{align}
  \phi_{\rm S}
  & = \int f_{\rm S}(t) \left[ \bm{k}_{\rm S} \cdot \big( \bm{a}_{\rm S} + \bm{a}_{\rm vib}(t) \big) - \alpha_{\rm S} \right] \dd t + \phi^{\rm sys}_{\rm S}, \\
  & = \phi_{\rm S}^{\rm acc} + \phi_{\rm S}^{\rm las} + \phi_{\rm S}^{\rm vib} + \phi_{\rm S}^{\rm sys},
\end{align}
\end{subequations}
where $S$ labels the atomic species, $f_{\rm S}(t)$ is the interferometer response function\cite{Cheinet2008}, $\bm{a}_{\rm S}$ is the acceleration due to gravity, $\bm{a}_{\rm vib}(t)$ is the acceleration of the common reference frame due to parasitic vibrations, $\alpha_{\rm S}$ is the chirp rate of one Raman frequency relative to the other. The integral of the response function gives the effective interrogation time, $T_{\rm eff}$, which for square Raman pulses yields \Eq \refeq{Teff2}.

For any high-sensitivity atomic gravimeter, vibrations of the reference frame are a major concern. In laboratories, these vibrations are strongly attenuated by using anti-vibration platforms. However, this solution is not compatible with mobile experiments such as ours\cite{Geiger2011, Barrett2016}, and our approach has been to correct for these vibrations by measuring them during the interferometer sequence using a three-axis MA (Nanometrics Titan) fixed to the back of the reference mirror. This force-balanced accelerometer has an auto-zeroing function that largely removes fixed offsets on each axis due to gravity. The MA signal projected along the Raman wavevector axis $\bm{k}_{\rm S} = k_{\rm S} \hat{\bm{z}}$ is
\be
  a_{\rm MA}(t) = \bm{c} \cdot \bm{a}_{\rm vib}(t) + b,
\ee
where $b$ is the MA bias, and $\bm{c} = (c_x,c_y,1+c_z)$ is a vector of coefficients, where each component $|c_i| \ll 1$. These account for differences in scale factors between the MA axes, and misalignments between the MA axes and $\bm{k}_{\rm S}$. We also note that the acceleration measured by the MA is intrinsically low-pass filtered due to the finite bandwidth of the device, but the low-frequency signals of interested are well within this bandwidth. Since both $\bm{c}$ and $b$ are temperature dependent, they can both vary during the experiment. However, we find that only $b$ varies significantly, and it does so slowly compared to the timescale of a single measurement ($\sim 3$ min). Since $\bm{a}_{\rm vib}(t)$ averages to zero by definition, we can remove most of the MA bias by subtracting the mean value of the MA signal during the measurement, $\langle a_{\rm MA} \rangle$. The resulting signal is integrated with the response function to obtain an estimate of the vibration-induced phase on each shot of the experiment
\be
  \label{tildephivib}
  \tilde{\phi}^{\rm vib}_{\rm S} = k_{\rm S} \int f_{\rm S}(t) \bm{c} \cdot \bm{a}_{\rm vib}(t) \dd t + k_{\rm S} \Delta b T_{\rm eff}^2
\ee
where $\Delta b = b - \langle a_{\rm MA} \rangle$ is a residual bias at the level of a few $\mu g$ resulting from our limited knowledge of the true one. This random phase is correlated with the output of the corresponding interferometer during each shot of the experiment. This way we are able to distinguish the part of the inertial phase due to gravity and the one from vibration noise, and hence we remove the contributions from the latter. Our measurement consists of scanning the chirp rate for each species and applying a post-correction $\tilde{\phi}^{\rm vib}_{\rm S}/T_{\rm eff}^2$ according to \Eq \refeq{tildealpha}. By fitting these reconstructed fringes, we extract $\alpha_{\rm S}^*$, which corresponds to the position of the central dark fringe. Finally, we measure the gravitational acceleration directly from $\alpha_{\rm S}^*/k_{\rm S}$.

However, this method is not perfect. The true vibration-induced phase differs from the estimate in \Eq \refeq{tildephivib} by the residual phase $\Delta\phi_{\rm S}^{\rm vib} = \phi_{\rm S}^{\rm vib} - \tilde{\phi}_{\rm S}^{\rm vib}$:
\be
  \label{phivibres}
  \Delta\phi_{\rm S}^{\rm vib} = k_{\rm S} \int f_{\rm S}(t) (\hat{\bm{z}} - \bm{c}) \cdot \bm{a}_{\rm vib}(t) \dd t - k_{\rm S} \Delta b T_{\rm eff}^2.
\ee
The first term is due to errors in the MA scale factor and its imperfect coupling with the mirror. This term appears as noise on the reconstructed fringes, and mainly limits the short-term sensitivity of each measurement. The second term is a bias due to the method. Although it averages to zero over many measurements, it limits our knowledge of $a_{\rm S}$ to the level at which we know the MA bias $b$. At the dark fringe total phase shift \refeq{phitot} is zero, hence we have
\begin{subequations}
\begin{align}
  \big(\alpha_{\rm S}^* - \bm{k}_{\rm S} \cdot \bm{a}_{\rm S}\big) T_{\rm eff}^2
  & = \Delta \phi_{\rm S}^{\rm vib} + \phi_{\rm S}^{\rm sys}, \\
  \frac{\alpha_{\rm S}^*}{k_{\rm S}} & = a_{\rm S} - \Delta b - \frac{\phi_{\rm S}^{\rm sys}}{k_{\rm S} T_{\rm eff}^2},
  \label{aMeas}
\end{align}
\end{subequations}
where $a_{\rm S} = \bm{k}_{\rm S} \cdot \bm{a}_{\rm S}/k_{\rm S}$ is the projection of the gravitational acceleration on $\bm{k}_{\rm S}$. The last equation shows that the measured acceleration has two main error contributions: a bias $\Delta b$ due to the MA, and one due to systematics effects. The measured E\"{o}tv\"{o}s parameter is then
\be
  \label{etaraw}
  \eta^{\rm raw}
  = \frac{1}{g} \left( \frac{\alpha_{\rm K}^*}{k_{\rm K}} - \frac{\alpha_{\rm Rb}^*}{k_{\rm Rb}} \right)
  = \frac{a_{\rm K} - a_{\rm Rb}}{g} + \eta^{\rm sys},
\ee
where $g = 9.805642$ m/s$^2$ is the known local gravitation acceleration, and $\eta^{\rm sys}$ contains all systematic bias terms. We note that the bias $\Delta b$ is common to both species by construction, hence it's contribution cancels in \Eq \refeq{etaraw}.

Finally, to evaluate $\eta^{\rm sys}$, we take into account the error due to a misalignment between $\bm{k}_{\rm Rb}$ and $\bm{k}_{\rm K}$. Assuming $\bm{a}_{\rm S} = a_{\rm S} \hat{\bm{z}}$ and defining $\bm{k}_{\rm K} = k_{\rm K} \hat{\bm{z}}$ and $\bm{k}_{\rm Rb} \cdot \hat{\bm{z}} = k_{\rm Rb} \cos \theta$, where $\theta$ is a small angle between the two wavevectors, we have
\be
  \frac{\bm{k}_{\rm K}}{k_{\rm K}} \cdot \bm{a}_{\rm K} - \frac{\bm{k}_{\rm Rb}}{k_{\rm Rb}} \cdot \bm{a}_{\rm Rb}
  \simeq a_{\rm K} - a_{\rm Rb} + \frac{\theta^2}{2} a_{\rm Rb}.
\ee
It follows from \Eq \refeq{aMeas} that $\eta^{\rm sys}$ contains three main terms
\be
  \eta^{\rm sys} = \frac{\theta^2}{2} \frac{a_{\rm Rb}}{g} + \frac{\phi_{\rm Rb}^{\rm sys}}{k_{\rm Rb} g T_{\rm eff}^2} - \frac{\phi_{\rm K}^{\rm sys}}{k_{\rm K} g T_{\rm eff}^2}.
\ee

\section{Bayesian estimation of the differential phase}
\label{sec:BayesianEstimation}


Bayesian estimation relies on Bayes' rule, which is a fundamental relationship between statistical probability distributions
\be
  \label{BayesRule}
  P(V|M) = \frac{p(V) L(M|V)}{N(M)}.
\ee
Here, $P(V|M)$ is the ``posterior'' probability which represents our knowledge of some variable of interest $V$, given a set of measurements $M$ of some system quantities. $p(V)$ is the ``prior'' probability before any measurements are made. $L(M|V)$ is the ``likelihood'' of obtaining the set $M$ \emph{given} $V$, and is computed based on a model of the system noise. Finally, $N(M) = \int L(M|V) p(V) \dd V$ is just a normalization factor for $P(V|M)$. The principle of Bayesian estimation is that the knowledge of $V$ can be improved on a measurement-by-measurement basis, with each successive measurement decreasing the uncertainty in the estimate of $V$. A well-known example of this type of recursive analysis is a Kalman filter \cite{Kalman1960, Cheiney2018}, which is a Bayesian estimator for the specific case of linear systems with white Gaussian noise. For the specific case of two coupled atom interferometers, the variable of interest is $\phi_{\rm d}^{\ud}$ and the $i^{\rm th}$ system measurement is given by the pair of normalized atomic sensor outputs $M_i = \{n_{\rm K}^{\ud}, n_{\rm Rb}^{\ud}\}_i$.

Any Bayesian estimator requires a good knowledge of the distribution of noise in the system, and the main challenge is to compute the likelihood distribution $L\big( \{n_{\rm K}^{\ud}, n_{\rm Rb}^{\ud}\} | \phi_{\rm d}^{\ud} \big)$ given a specific noise model for $n_{\rm K}^{\ud}$ and $n_{\rm Rb}^{\ud}$. To illustrate the possible noise sources in this system, we modify the definitions of the $n_{\rm S}^{\ud}$ in \Eqs \refeq{Lissajous}
\begin{subequations}
\label{LissajousNoise}
\begin{align}
  n_{\rm K}^{\ud}(\phi_{\rm c}) & = (1 + \epsilon_{A_{\rm K}})\cos(\kappa \phi_{\rm c} + \phi_{\rm d}^{\ud} + \epsilon_{\phi_{\rm d}}) + \epsilon_{n_{\rm K}}, \\
  n_{\rm Rb}^{\ud}(\phi_{\rm c}) & = (1 + \epsilon_{A_{\rm Rb}})\cos(\phi_{\rm c}) + \epsilon_{n_{\rm Rb}}.
\end{align}
\end{subequations}
The stochastic quantities $\epsilon_{A_{\rm S}}$, $\epsilon_{n_{\rm S}}$, and $\epsilon_{\phi_{\rm d}}$ represent uncorrelated noise in the single sensor amplitudes, offsets, and differential phase, respectively. Each of these stochastic terms is assumed to follow a Gaussian probability distribution with zero mean and standard deviation given by $\sigma_{A_{\rm S}}$, $\sigma_{n_{\rm S}}$, and $\sigma_{\phi_{\rm d}}$, respectively. In our system, the measured amplitude noise ($\sigma_{A_{\rm S}} = \sigma_{C_{\rm S}}/2$) is small compared to other sources (see Table \ref{tab:FringeParameters}), hence we neglect this term in the noise model. The remaining terms arise due to detection noise, which affects the interferometer offsets, and non-common phase noise from systematic effects which directly affect the differential phase.

\section{Estimating the noise parameters of coupled atom interferometers}
\label{sec:NoiseEstimation}


In this appendix, we describe our technique for estimating the noise parameters of two coupled atom interferometers. Depending on the analysis method (FRAC or Bayesian), we take two separate approaches.

For the FRAC method, we use the fact that each interference fringe is observable and follows the simple model
\be
  \label{yModel}
  y(\phi) = Y - \frac{C}{2} \cos(\phi - \Phi).
\ee
It is straightforward to estimate the set of fringe parameters $F = \{Y, C, \Phi\}$ from fits to the data, as shown in \Fig \ref{fig:Fringes}. Each of these quantities is assumed to follow a Gaussian probability distribution, with mean values given by the measured fit parameters. We wish to estimate the standard deviations associated with the offset, contrast and phase: $\sigma_Y$, $\sigma_C$ and $\sigma_\Phi$. This can be done using a maximum-likelihood approach \cite{VanderPlas2014, Cheiney2018}. Given a single atomic measurement $D_i = \{\phi_i, y_i\}$, the probability of measuring $D_i$ given the aforementioned fringe parameters is
\be 
  p(D_i|F) = \frac{1}{\sqrt{2\pi \sigma^2}} \exp\left( -\frac{\big(y_i - y(\phi_i)\big)^2}{2\sigma^2} \right),
\ee
where $\sigma$ is the standard deviation of fit residuals, which is a statistical mixture of all possible noise sources. Assuming there are no correlations between parameters, from \Eq \refeq{yModel} we have
\be 
  \sigma^2 = \sigma_Y^2 + \Big(\frac{\sigma_C}{2}\cos(\phi - \Phi)\Big)^2 + \Big( \frac{C}{2}\sin(\phi - \Phi) \sigma_\Phi \Big)^2.
\ee
The likelihood distribution is simply the product of the probabilities for $N$ measurements:
\be 
  L(D|F) = \prod_{i=1}^N p(D_i|F).
\ee
To obtain estimates of the noise parameters, we maximize this likelihood given a set of measurements $D = \{D_i\}$. Since the likelihood distribution is a product of Gaussian functions, it is convenient to instead minimize the negative logarithm:
\be
  -\log L(D|F) = \frac{N}{2} \log\big(2\pi\sigma^2\big) + \sum_{i=1}^N \frac{\big(y_i - y(\phi_i)\big)^2}{2\sigma^2}.
\ee
We employ the robust Nelder-Mead multidimensional minimization routine to simultaneously estimate $\sigma_Y$, $\sigma_C$ and $\sigma_\Phi$.

For the Bayesian analysis, we assume each interference fringe is washed out by phase noise and cannot be observed directly. Instead, to estimate the (normalized) offset noise parameters $\sigma_{n_{\rm K}}$ and $\sigma_{n_{\rm Rb}}$, we use the mean uncertainty in the probabilities $P_S^{\ket{2}}$ derived from a given set of detection traces (see \Fig \ref{fig:Detection}). We also add a contribution to this noise from the operation of normalizing the sensor outputs due to uncertainties in the fringe offset $Y_S$ and contrast $C_S$ (which are estimated separately based on the statistical distribution of $P_S^{\ket{2}}$, as in \Ref \citenum{Geiger2011}). The differential phase noise $\sigma_{\phi_d}$ is more challenging, because it cannot be estimated directly from the interferometer outputs. We use an iterative approach to estimate this parameter. First, we run the algorithm on a small sample of data using an initial estimate of $\sigma_{\phi_d}$ to obtain an initial estimate of $\phi_d$. Then we re-insert this fixed value of $\phi_d$ into the algorithm and use it to estimate $\sigma_{\phi_d}$. We iterate between these two estimates until we find convergence. We also verify that the standard deviation of the resulting distribution of $\phi_d$ estimates is consistent with the individual uncertainties returned by the algorithm. 



\bibliography{References}

\end{document}